\documentclass[fleqn,usenatbib,useAMS]{mnras}

\usepackage{graphicx}	
\usepackage{amsmath}	
\usepackage{amssymb}	
\usepackage{multicol}        
\usepackage{bm}		
\usepackage{pdflscape}	

\usepackage{xcolor}
\usepackage{color}
\usepackage{graphicx}
\usepackage{dcolumn}
\usepackage{bm}
\usepackage{hyperref}
\usepackage{amsmath}
\usepackage{multirow}
\usepackage{cancel}

\newcommand{\Ms}{{\rm ~M}_\odot}

\newcommand{\gw}{{\rm GW}}

\newcommand{\ham}{{\rm HM}}
\newcommand{\kms}{{\rm ~{\rm km}/{\rm s}}}
\newcommand{\nbsix}{\textsc{Nbody6++GPU~}}

\newcommand{\bse}{\textsc{BSE~}}
\newcommand{\mcl}{\textsc{McLuster~}}

\newcommand{\dragonii}{\textsc{Dragon-II~}}

\usepackage[T1]{fontenc}
\usepackage{ae,aecompl}

\title[The \textsc{Dragon-II} simulations - II.]{The \textsc{Dragon-II} simulations - II. Formation mechanisms, mass, and spin of intermediate-mass black holes in star clusters with up to 1 million stars}

\author[M. Arca Sedda et al]{Manuel Arca Sedda$^{1,2,3}$%
\thanks{Contact e-mail:\href{mailto:m.arcasedda@gmail.com}{manuel.arcasedda@gssi.it}},
Albrecht W. H. Kamlah$^{4,3}$, Rainer Spurzem$^{3,5,6}$,
\newauthor
Francesco Paolo Rizzuto$^{7}$, Mirek Giersz$^{9}$, Thorsten Naab$^{7}$, Peter Berczik$^{3,10,11}$
\\
$^{1}$ Gran Sasso Science Institute (GSSI), 67100 L’Aquila, Italy\\
$^{2}$ Physics and Astronomy Department Galileo Galilei, University of Padova, Vicolo dell'Osservatorio 3, I--35122, Padova, Italy\\
$^{3}$ Astronomisches Rechen-Institut, Zentrum f\"{u}r Astronomie der Universit\"{a}t  Heidelberg, M\"onchhofstr. 12-14, D-69120 Heidelberg, Germany\\
$^{4}$ Max-Planck-Institut f\"ur Astronomie, K\"onigstuhl 17, 69117 Heidelberg, Germany \\
$^{5}$ National Astronomical Observatories and Key Laboratory of Computational Astrophysics, Chinese Academy of Sciences, \\ 20A Datun Rd.,Chaoyang District, 100101, Beijing, China\\
$^{6}$ Kavli Institute for Astronomy and Astrophysics, Peking University, Yiheyuan Lu 5, Haidian Qu, 100871, Beijing, China\\
$^{7}$ Department of Physics, University of Helsinki, Gustaf Hällströmin katu 2, FI-00014, Helsinki, Finland\\
$^{8}$ Max Planck Institute for Astrophysics, Karl-Schwarzschild-Str. 1, D-85740, Garching, Germany\\
$^{9}$ Nicolaus Copernicus Astronomical Centre, Polish Academy of Sciences, ul. Bartycka 18, 00-716 Warsaw, Poland\\
$^{10}$ Konkoly Observatory, Research Centre for Astronomy and Earth Sciences, E\"otv\"os Lor\'and Research Network (ELKH), \\ MTA Centre of Excellence, Konkoly Thege Mikl\'os \'ut 15-17, 1121 Budapest, Hungary \\
$^{11}$ Main Astronomical Observatory, National Academy of Sciences of Ukraine, 27 Akademika Zabolotnoho St., 03680, Kyiv, Ukraine \\
}

\date{\today}

\pubyear{2022}

\begin{document}
\label{firstpage}
\pagerange{\pageref{firstpage}--\pageref{lastpage}}
\maketitle

\begin{abstract}
The processes that govern the formation of intermediate-mass black holes (IMBHs) in dense stellar clusters are still unclear. Here, we discuss the role of stellar mergers, star-BH interactions and accretion, as well as BH binary (BBH) mergers in seeding and growing IMBHs in the \textsc{Dragon-II} simulation database, a suite of 19 direct $N$-body models representing dense clusters with up to $10^6$ stars. \textsc{Dragon-II} IMBHs have typical masses of $m_{\rm IMBH} = (100-380)\Ms$ and relatively large spins $\chi_{\rm IMBH} > 0.6$. We find a link between the IMBH formation mechanism and the cluster structure. In clusters denser than $3\times 10^5$ M$_\odot$ pc$^{-3}$, the collapse of massive star collision products represents the dominant IMBH formation process, leading to the formation of heavy IMBHs ($m_{\rm IMBH} > 200$ M$_\odot$), possibly slowly rotating, that form over times $<5$ Myr and grow further via stellar accretion and mergers in just $<30$ Myr. BBH mergers are the dominant IMBH formation channel in less dense clusters, for which we find that the looser the cluster, the longer the formation time ($10-300$ Myr) and the larger the IMBH mass, although remaining within $200$ M$_\odot$. Strong dynamical scatterings and relativistic recoil efficiently eject all IMBHs in \textsc{Dragon-II} clusters, suggesting that IMBHs in this type of cluster are unlikely to grow beyond a few $10^2$ M$_\odot$.
\end{abstract}




\section{Introduction}

Despite the great progresses in observations, marked by the detection of intermediate-mass black hole (IMBH) candidates with masses as low as $50,000\Ms$ \citep{2018ApJ...863....1C}, and the first detection of an IMBH with mass $\sim 150 \Ms$  formed from the merger of two massive stellar BHs \citep[named GW190521][]{2020PhRvL.125j1102A}, IMBHs remain elusive objects whose existence in the $M_{\rm IMBH} = 10^2-10^5\Ms$ mass range is largely debated \cite[see recent reviews by][]{2017IJMPD..2630021M,2020ARA&A..58..257G}. 

Several IMBH candidates have been proposed in galactic and extragalactic clusters \citep{2013ApJ...769..107L,2013A&A...552A..49L,2008MNRAS.389..379M,2019MNRAS.482.4713Z,2016IAUS..312..197Z,2018anms.conf...15W,2010ApJ...710.1063V,2012ApJ...750L..27S,2018NatAs...2..656L,2017Natur.542..203K,2008AJ....135..182B,2017MNRAS.464.3090A,2016MNRAS.455...35A,2018MNRAS.481..627A,2022A&A...661A..68T,2019arXiv190500902A}, but none of the explorations conducted so far led to conclusive results, making IMBH formation processes one of the most intriguing puzzles of modern astronomy.

Numerical and theoretical works on IMBH formation in dense star clusters suggest that the IMBH seeding can occur via three, rather uncertain, pathways \citep{2002ApJ...576..899P,2004Natur.428..724P,2015MNRAS.454.3150G,2016MNRAS.461..948M,2016ApJ...831..187A,2018ApJ...856...92F,2019MNRAS.486.5008A,2020ApJ...894..133A,2021MNRAS.507.5132D,2021ApJ...908L..29G,2021MNRAS.501.5257R,2022MNRAS.512..884R,2021ApJ...920..128A,2021A&A...652A..54A}: multiple stellar mergers, accretion of stellar matter onto a stellar BH, or repeated BH mergers. These mechanisms are not mutually exclusive: multiple stellar mergers can form a very massive star (VMS) that eventually collides with a stellar BH and the collision product further grows by merging with other BHs in the cluster. These processes could explain the formation of supemassive BHs (SMBHs) in galactic nuclei \citep{1984ARA&A..22..471R}. A further formation channel could be via formation and collapse of a supermassive star, the so-called direct collapse scenario for SMBH seedings in galactic nuclei \citep{1998A&A...331L...1S,2001ApJ...551L..27M,2003ApJ...596...34B,2009MNRAS.396..343R}. A similar process, aided by stellar collisions and gaseous accretion, could operate also in the most massive globular clusters, provided that they accrete a significant amount of the gas in which they are embedded at formation \citep{2018MNRAS.478.2461G}.

The impact of multiple stellar mergers onto the IMBH buildup depends in part on the possible insurgence of pair-instability (PISN) and pulsational pair-instability supernova (PPISN) mechanisms. Stars that develop an He core with mass in the range $m_{\rm He}=(64-135)\Ms$  undergo PISN and explode leaving no remnant, whilst stars $m_{\rm He}=(32-64)\Ms$ suffer a strong mass loss owing to PPISN and leave remnants with a mass generally lighter than $40-50\Ms$. These explosive mechanisms result in the so-called upper mass-gap, a region of the mass spectrum $m_{\rm BH} = 40-150\Ms$ where no BHs are expected. The boundaries of the upper mass-gap are rather uncertain, and depend on many details, among which the stellar evolution model, stellar rotation, the rate of thermonuclear reactions \citep{2021ApJ...912L..31W,2021MNRAS.504..146V,2019ApJ...882..121S,2019ApJ...887...53F,2021MNRAS.501.4514C,2021MNRAS.502L..40F,2021MNRAS.505.2170T,2020ApJ...888...76M}. Stellar mergers can actually overcome PISN and PPISN by mixing stars in different evolutionary stage, a mechanism that permits to increase the stellar mass but keep the He core below the threshold for these explosive mechanisms to develop \citep[see e.g.][]{2019MNRAS.485..889S}. Stellar mergers of this type have proven to be a viable way to generate upper-mass gap BHs in star clusters and, in some cases, IMBHs \citep{2021MNRAS.507.5132D, 2020MNRAS.497.1043D, 2020ApJ...903...45K,2021MNRAS.501.5257R,2021ApJ...920..128A,2022MNRAS.512..884R, 2021ApJ...908L..29G,2022A&A...665A..20B,2022MNRAS.516.1072C,2022ApJS..258...22R,2022arXiv220403493B, 2022MNRAS.515.5106W}.

Whilst there is some general consensous about the outcome of stellar mergers, also thanks to the development of detailed hydrodynamical simulations coupled with stellar evolution models \citep{2022arXiv220403493B,2022MNRAS.516.1072C}, it is still rather unclear how much mass a massive star can accrete onto a stellar BH. Several works have shown that in the case of a "normal" star merging with a stellar BHs, there is little accretion as most of the energy is radiated away via jets, although the mechanism is highly uncertain and likely depends on the star structure and evolutionary stage \citep{2013ApJ...767...25G,2015ApJ...798L..19M, 2020ApJ...894..147C,2022ApJ...933..203K}. Hydrodynamical simulations of star-BH close interactions have shown that up to $70\%$ of the star mass remains bound to the BH, but energy arguments suggest that even a tiny amount of accreted matter, $O(10^{-3}-10^{-2}\Ms)$, generates enough energy to evaporate the accretion disk and halt the BH growth \citep{2022ApJ...933..203K}. Nonetheless, recent simulations modelling the common envelope phase of a tight star-BH binary have shown that the BH accretes the stellar core and expels the envelope, a process -- possibly accompanied by a SN-like transient -- that can spin-up the BH to nearly extremal values regardless the initial spin \citep{2020ApJ...892...13S}. In multiple main sequence (MS) star collisions, the merger product is expected to be characterised by a compact core and a tenuous envelope with densities as low as $10^{-10}$ g cm$^{-3}$ \citep{2009A&A...497..255G}. Therefore, it seems reasonable to assume that a BH would eat-up a significant fraction of mass from a massive companion that underwent multiple stellar mergers. Given this, recent works parametrised the amount of accreted matter through an accretion parameter $f_c=0-1$ \citep{2017MNRAS.467..524B, 2021MNRAS.501.5257R,2021ApJ...920..128A, 2022MNRAS.512..884R, 2022arXiv221113320R}. 

Repeated BH mergers can potentially build-up upper-mass gap BHs and IMBHs, but their efficiency is inevitably hampered by the development of post-merger recoil originated by anysotropic GW emission \citep[e.g.][]{2007PhRvL..98w1102C, 2008PhRvD..77d4028L, 2012PhRvD..85h4015L}, which can easily eject the post-merger product from the parent environment, especially in star clusters with velocity dispersion $\sigma < 100$ km s$^{-1}$ \citep{2008ApJ...686..829H,2022ApJ...927..231F,2018ApJ...856...92F,2021A&A...652A..54A,2021ApJ...920..128A,2020ApJ...894..133A,2021arXiv210912119A,2021ApJ...918L..31M}. 
Typically, the amplitude of the kick imparted promptly after a merger on the remnant depends on the binary mass ratio and the amplitude and direction of the component spins, and can attain values that span more than two orders of magnitude.
Despite its crucial impact on post-merger dynamics, little is known about the natal spin of stellar BHs, let alone IMBHs. Observations of several high-mass X-ray binaries show that BHs in these systems are nearly maximally spinning  \citep[see e.g.][]{2019ApJ...870L..18Q,reynolds21}, while observations of GW sources suggest that merging BHs are mostly slowly rotating ($\chi_{\rm BH} < 0.5$) \citep{2021arXiv211103634T}. 
From the theoretical point of view, it has been suggested that the evolution of the BH stellar progenitors could significantly impact the natal spin distribution. 
In single stars and binaries with negligible mass transfer, efficient angular momentum transport driven by magnetic fields could trigger the formation of BHs with natal spins as small as $\chi_{\rm BH} \lesssim 0.01$ via Taylor-Spruit dynamo \citep{2019ApJ...881L...1F}. 
A significant mass-transfer can, instead, significantly spin-up a BH even if it is spinless at birth, possibly explaining the observed spin of BHs in Galactic low-mass X-ray binaries ($\chi_{\rm BH} \sim 0.1-0.99$) \citep{2015ApJ...800...17F}. Similarly, accretion from a BH progenitor onto a close companion in a binary and subsequent accretion from the companion onto the BH can spin-up the BH in high-mass X-ray binaries, provided that the angular momentum transfer when the companion leaves the MS phase is inefficient \citep{2019ApJ...870L..18Q,2022ApJ...938L..19G}. High-mass X-ray binaries with highly spinning BHs are not expected to produce merging BHs, a feature that partly explains the dearth of highly spinning BHs in observed BH mergers \citep{2022ApJ...938L..19G}. 
In massive binaries undergoing both Roche lobe overflow and common envelope and eventually forming a BH binary (BBH), the first-born BH can have nearly zero spin or a spin covering a wide range, depending on the stellar prescription adopted, whilst the second BH could have nearly extremal spin \citep{2018A&A...616A..28Q, 2020A&A...635A..97B, 2020A&A...636A.104B}. This is likely driven by tidal synchronization of BH progenitors rotation and their mutual orbit \citep{2016MNRAS.462..844K, 2017ApJ...842..111H}. Nonetheless, massive binaries could also form BHs with negligible spins, provided that their progenitors lose their hydrogen envelope before undergoing SN \citep{2020A&A...635A..97B, 2020A&A...636A.104B}.
In the case of BHs formed from star-BH mergers, instead, it has been shown that the accretion of the star core onto the BH can spin-up the BH to extreme values \citep{2020ApJ...892...13S}. The aforementioned scenarios for BH natal spin can have a significant impact on the properties of IMBHs, depending on their formation mechanism. An IMBH formed via star-BH merger, for example, could be characterised by a large spin, while one formed via the collapse of a VMS could have negligible spin. 

Stellar mergers, star-BH interactions, and BBH mergers can also have an impact on the formation of BHs in the upper-mass gap. In the first three observation runs, the LIGO-Virgo-Kagra collaboration (LVC) revolutionized our knowledge of BHs, proving the existence of BHs in and beyond the upper-mass gap. The most updated GW transient catalog (GWTC-3) contains 85 sources associated with the merger of two BHs with a mass above $m_{\rm BH} = 3\Ms$ \citep{2021arXiv211103634T,2021arXiv211103606T}. Around one-third of them (27) have one component above $m_{\rm BH} > 40.5\Ms$, and 8 of them have one component heavier than $m_{\rm BH} > 65\Ms$, i.e. two proposed lower limits for the PISN \citep{2016A&A...594A..97B,2017MNRAS.470.4739S}. Moreover, 8 sources have a remnant mass $m_{\rm BH, rem} > 100\Ms$, 3 of which exceeds the IMBH threshold at 95 confidence level. With the forthcoming fourth observation run (O4), the LVC collaboration will possibly detect further 30-150 merging events, thus future detection will provide further insights on the development of BH mergers with upper-mass gap BHs. 

In this work, we discuss the formation of IMBHs and upper mass-gap BHs in the \dragonii star cluster database, a suite of 19 direct $N$-body simulations of star clusters comprised of up to 1 million stars and up to $33\%$ of stars initially in binaries (details about these models are discussed in our companion paper, Arca Sedda et al in prep), performed with the \nbsix code\footnote{\url{https://github.com/nbody6ppgpu/Nbody6PPGPU-beijing}} \citep{2015MNRAS.450.4070W,2022MNRAS.511.4060K}. 

The paper is organised as follows: in Section \ref{sec:methods} we briefly summarise the main features of our models; Section \ref{sec:imbh} describes how IMBHs form in \dragonii simulations and what is the impact of different formation channels; whilst Section \ref{sec:disc} is devoted to discuss the impact of Newtonian and relativistic dynamics on the mass and spin of IMBHs in dense star clusters. Section \ref{sec:end} summarises the main results of the work.

\section{Numerical Methods}
\label{sec:methods}

\subsection{Modelling \dragonii clusters with the \nbsix code}

All \dragonii clusters are represented by \cite{1966AJ.....71...64K} models with an dimensionless potential well $W_0 = 6$, a number of stars of $N = (120 - 300 - 600)\times 10^3$, and an initial half-mass radius either $R_\ham = 0.47, ~0.80, ~1.75$ pc. As described in the first paper of the series (Arca Sedda et al, subm., hereafter paper AS-I) this choice is compatible with observations of several Galactic young massive clusters and produce cluster models that broadly match observed masses and half-mass radii of dense clusters in the Magellanic clouds (see Figure 2 in paper AS-I). For all models we adopt a binary fraction $f_b=0.2$\footnote{Note that the binary fraction is defined as $f_b = n_b/(n_s+n_b)$, where $n_b$ is the number of binaries. This implies that the fraction of stars initially in binary systems is $f_{2b} = 2f_b/(1+f_b)= 0.10-0.33$, with $f_b=0.05, 0.2$.}, defined as the number of binaries normalised to the sum of the number of single stars and binary pairs. For models with $R_\ham = 2.2$ pc, we run an additional series of models where we adopt $f_b = 0.05$ and $N = (120 - 300 - 1,000)\times 10^3$. All clusters have the same metallicity, $Z = 0.0005$, a value consistent with the metallicity of several globular clusters in the Milky Way that may host a substantial population of BHs \citep{2018MNRAS.479.4652A,2018MNRAS.478.1844A,2020ApJ...898..162W}.
The reduced computational cost of modelling a smaller amount of binaries permitted us to increase the total number of stars to one million, which is the maximum amount of stars and binaries ever simulated for realistic star cluster models with a direct $N$-body code \citep{2016MNRAS.458.1450W}. 

All clusters have been initialised with the \mcl code \cite[][, Leveque in prep]{2011MNRAS.417.2300K, 2022MNRAS.511.4060K, 2022MNRAS.514.5739L}, adopting a \citep{2001MNRAS.322..231K} initial mass function limited between $0.08\Ms$ and $150\Ms$. Binary eccentricities are drawn from a thermal distribution, whilst semimajor axes follow a distribution flat in logarithmic values limited between the sum of stellar radii and 50 AU \citep{2015MNRAS.450.4070W, 2022MNRAS.511.4060K}. Binary components are paired according to a uniform mass ratio distribution if their mass exceeds $m_*>5\Ms$, whilst lighter stars are paired randomly \citep{2012ApJ...751....4K,2012Sci...337..444S,2014ApJS..213...34K}. 
All clusters are assumed to orbit on a circular orbit 13.3 kpc away from the centre of a galaxy with total mass $1.78\times10^{11}\Ms$, assuming for the galaxy a Keplerian gravitational potential. Note that the choice of parameters is such that the velocity curve at the adopted distance is similar to the one observed in the Milky Way. This implies that all \dragonii clusters are initially well contained inside their Roche lobe, thus the galactic field has little effect on the cluster structural evolution. In all cases but one, we ran two different realisation of each cluster to reduce the impact of statistical fluctuations. 
Table \ref{tab:t1} summarizes the main properties of \dragonii clusters. The table shows the initial parameters of the clusters, the simulated time $T_{\rm sim}$, the number of merging compact objects occurring inside the cluster or after their ejection, the absolute maximum mass attained by BHs and the maximum BH mass at the end of the simulation, the number of BHs with a mass above $30 \Ms$ or $40\Ms$. For each set of initial conditions, we provide numbers for each independent realisation.

The simulations have been performed with the \nbsix code, a state-of-the-art direct $N$-body integrator that exploits GPU-accelerated high-performance supercomputing \citep{1999JCoAM.109..407S,2012MNRAS.424..545N,2015MNRAS.450.4070W,2022MNRAS.511.4060K}. The current version of the code follows the footstep of a 50 year old tradition initiated by Sverre Aarseth \citep{1974A&A....37..183A,1999JCoAM.109..407S,1999PASP..111.1333A,2003gnbs.book.....A,2008LNP...760.....A,2012MNRAS.424..545N,2015MNRAS.450.4070W,2022MNRAS.511.4060K}. 

The code exploits a 4th-order Hermite integrator with individual block-time step \citep{1986LNP...267..156M,1995ApJ...443L..93H} and implements a dedicated treatment for close encounters and few-body dynamics based on the Kustaanheimo-Stiefel (KS) regularisation \citep{Stiefel1965}, the Ahmad-Cohen (AC) scheme for neighbours \citep{1973JCoPh..12..389A}, and algorithmic chain regularisation \citep{1999MNRAS.310..745M,2008AJ....135.2398M}, which permits to resolve the evolution of binaries with a period $10^{-10}$ times smaller than the typical dynamical timescales of star clusters. 

Recently, a series of improvements have been introduced in the code to treat the formation and merger of relativistic binaries \citep{2021MNRAS.501.5257R} and quantify the fraction of stellar matter that can be fed to a stellar BH in binary systems or star-BH collisions \citep{2022MNRAS.512..884R}.
Stars in \dragonii clusters are evolved self-consistently from the zero age main sequence through the \bse code \citep{2002MNRAS.329..897H}, conveniently updated to feature state-of-the-art recipes for the evolution of massive stars, the mass spectrum and natal kicks of BHs and NSs, and the physics of (P)PISN, \citep[for a detailed description of stellar evolution in \nbsix, see][]{2020A&A...639A..41B,2021MNRAS.500.3002B,2022MNRAS.511.4060K}. In this work, we use the so-called level-B of stellar evolution \citep[][Arca Sedda et al, in prep]{2022MNRAS.511.4060K}.

After a series of major upgrades described in recent papers \citep[][Arca Sedda et al, in prep]{2021MNRAS.501.5257R, 2022MNRAS.511.4060K, 2022MNRAS.512..884R}, \nbsix currently implements multiple choices for the distributions of BH natal spins and numerical relativity fitting formulae to calculate the final mass and spin of merger remnants, based on \citep{2017PhRvD..95f4024J}, and the relativistic recoil imparted onto them because of asymmetric GW emission, based on \citep{2007PhRvL..98w1102C,2008PhRvD..77d4028L,2012PhRvD..85h4015L}. 

Although \nbsix implements the GW recoil in a self-consistent way, the amplitude of the recoil depends primarily on the merging masses and the spin amplitude and orientation, making the process highly stochastic. Given the relatively small number of simulations in our sample, we decide to explore the role of post-merger kicks as follows.

Firstly, we run all simulations assuming zero GW recoil.
Secondly, we calculate the typical GW recoil experienced by merger products in \dragonii clusters and infer the corresponding retention probability in post-process, following an approach widely used in the literature. 
Thirdly, in case of a simulation featuring multiple generation mergers, we re-run the simulation shortly before the $n$-th merger with the GW kicks enabled to verify if, upon retention, the BH undergoes an $n+1$-th generation merger. 

The scopes of such simplified scheme are manifold. On the one hand, it permits us to verify whether multiple-generation mergers can occur in absence of relativistic effects. On the other hand, it permits us to assess the impact of Newtonian and general relativistic dynamics on the formation and retention of IMBHs. Furthermore, using this multi-stepped procedure helps us to optimise the available computational resources and to maximise the scientific outputs of the simulations. 

\begin{table*}
    \centering
    \begin{tabular}{ccccccc|cc|cc|cc|cc|cc|cc|cc}
    \hline
    \hline
        $N_*$ & $M_c$ & $R_h$ & $f_b$ &  $N_{\rm sim}$ & $T_{\rm rlx}$ &$T_{\rm seg}$ & \multicolumn{2}{|c}{$T_{\rm sim}$} & \multicolumn{2}{|c}{$N_{\rm GW, in}$}& \multicolumn{2}{|c}{$N_{\rm GW, out}$} & \multicolumn{2}{|c}{$M_{\rm max}$} & \multicolumn{2}{|c}{$M_{\rm max,fin}$} & \multicolumn{2}{|c}{$N_{>30}$} & \multicolumn{2}{|c}{$N_{>40}$} \\
        $10^3$ & $10^5\Ms$ & pc & &  & Myr & Myr & \multicolumn{2}{|c}{Myr} & \multicolumn{2}{|c}{}&\multicolumn{2}{|c}{}& \multicolumn{2}{|c}{$\Ms$} & \multicolumn{2}{|c}{$\Ms$}& \multicolumn{2}{|c}{}& \multicolumn{2}{|c}{}\\
    \hline
    120 & 0.7 & 1.75 & 0.05& 2& 99 & 2.1 & 2379 & 2326 & 0 & 2 & 2& 0&  64 &   76 & 25 &  34 &   0 &   2 &  0 &  0 \\
    300 & 1.8 & 1.75 & 0.05& 2& 142 & 2.7 & 1196 & 1422 & 0 & 2 & 2& 2&  69 &   77 & 40 &  40 &  13 &  13 &  5 &  1 \\
    1000& 5.9 & 1.75 & 0.05& 2& 233 & 4.5 &  207 &  194 & 1 & 1 & 4& 4&  81 &  146 & 52 &  70 & 149 & 169 & 72 & 85 \\
    120 & 0.7 & 1.75 & 0.2 & 2& 99 & 2.1 & 1710 & 1540 & 2 & 2 & 0& 2& 232 &   81 & 38 &  28 &   2 &   0 &  0 &  0 \\
    300 & 1.7 & 1.75 & 0.2 & 2& 142 & 2.7 &  519 &  793 & 1 & 0 & 7& 5&  92 &   77 & 65 &  47 &  26 &  26 &  8 & 14 \\
    600 & 3.5 & 1.75 & 0.2 & 2& 189 & 3.4 &  205 &  126 & 0 & 0 & 2& 5&  87 &  144 & 59 &  84 &  95 & 103 & 45 & 65 \\
    120 & 0.7 & 0.80 & 0.2 & 2& 30 & 0.7 & 1154 & 1201 & 4 & 3 & 4& 2& 120 &  132 & 21 &  27 &   0 &   0 &  0 &  0 \\
    300 & 1.7 & 0.80 & 0.2 & 2& 44 & 0.8 &  307 &  309 & 1 & 0 & 1& 0&  93 &  107 & 40 &  43 &  15 &  11 &  2 &  2 \\
    120 & 0.7 & 0.47 & 0.2 & 2& 14 & 0.3 & 1149 &  530 & 2 & 2 & 3& 1& 350 &   92 & 50 &  30 &   1 &   0 &  1 &  0 \\
    300 & 1.7 & 0.47 & 0.2 & 1& 20 & 0.4 &  148 &    - & 4 & - & 3& -& 245 &    - & 48 &   - &  22 &   - &  9 &  - \\
    \hline
    \end{tabular}
    \caption{Col. 1-4: initial number of stars, cluster mass and half-mass radius, primordial binary fraction. Col. 5: number of indipendent realisations. Col. 6-7: initial relaxation and segregation time. Col. 8: simulated time. Col. 9-10: number of mergers inside the cluster. Col. 11: maximum BH mass during the simulation. Col. 12: maximum BH mass at the end of the simulation. Col. 13-14: number of BHs with a mass $m_{\rm BH}>30\Ms$ or $>40\Ms$ at the last simulation snapshot.}
    \label{tab:t1}
\end{table*}

\section{Intermediate-mass and upper-mass gap black holes formation in massive dense clusters}
\label{sec:imbh}

Out of 19 simulated clusters, we find 8 IMBHs with a mass $M_{\rm IMBH} = (107-350)\Ms$, corresponding to a formation probability of $P_{\rm IMBH}\sim 42\pm15\%$. Despite the small statistics, we note that there is a moderate dependence on the binary fraction and the cluster compactness. In fact, we find an IMBH formation fraction of $f_{\rm IMBH} = 0.17, ~0.33, ~0.75, ~0.67$ going from $f_b=0.05$ to $f_b=0.2$ and from $R_\ham = 1.75,~0.8,~0.47$ pc. Comparing different models makes evident the importance of binaries and cluster compactness in determining the IMBH seeding. 
The formation history and main properties of all IMBHs in \dragonii simulations are described in detail in Appendix \ref{sec:IMevo}.

Aside IMBHs, around $N_{\rm gap}\simeq 10^2$ upper mass-gap BHs form within the simulation time, corresponding to a formation efficiency 
\begin{equation}
    \eta_{\rm gap} = \frac{N_{\rm gap}}{M_{\rm sim}} = 3.44 \times 10^{-5}\Ms^{-1}, \\
\end{equation}
where $M_{\rm sim} = 3.65\times 10^6\Ms$ is the total simulated mass.

The formation of IMBHs and upper-mass gap BHs via stellar mergers, accretion of stellar material onto a stellar BH, BH-BH mergers, or a combination of them, intrinsically depend on the host cluster properties. The development of one mechanism or another is intrinsically linked to the initial cluster structure, which determines the typical timescales of dynamical processes. The earliest process that regulates the evolution of a star cluster with a broad mass spectrum is mass-segregation, by which the most massive stars sink toward the cluster centre and start dominating dynamics in the inner core \citep{1971ApJ...164..399S,1987degc.book.....S}. The mass-segregation timescale of heavy stars with maximum mass $m_{\rm max}$ can be expressed as \citep[e.g.][]{1971ApJ...164..399S,2004Natur.428..724P,2008gady.book.....B,2021MNRAS.501.5257R}
\begin{equation}
    T_{\rm seg} \sim \frac{0.138N\langle m_* \rangle}{m_{\rm max}\ln\left(0.11M_{cl}/m_{\rm max}\right)} \left(\frac{R_\ham^3}{GM_{cl}}\right)^{1/2}.
\end{equation}
If the mass-segregation time is shorter than the lifetime of the most massive stars, it implies that they will sink to the centre before they turn into compact objects, thus their interactions can trigger more easily stellar collisions or massive star-BH close interactions. 
As summarised in Table \ref{tab:t1}, \dragonii clusters have a typical mass-segregation time $T_{\rm seg}=0.4-3.4$ Myr, thus they represent ideal laboratories to study the impact of star mergers and strong interactions on the early evolution of star clusters. 

In the following section, we describe the impact of stellar collisions, star-BH collisions and mergers, and compact object mergers on the formation of IMBHs and mass-gap BHs. 

\subsection{Formation channels and formation times}

Despite the relatively small database, our \dragonii models support the formation of IMBHs via all the three main channels, complementing previous works \citep{2002ApJ...576..899P,2004Natur.428..724P,2015MNRAS.454.3150G,2021MNRAS.501.5257R,2021ApJ...908L..29G,2022MNRAS.512..884R,2022MNRAS.514.5879M}.

To provide the reader with a clearer idea about how IMBHs form in \dragonii clusters, we provide below two examples extracted from our simulations.

\begin{figure}
    \centering
    \includegraphics[width=0.9\columnwidth]{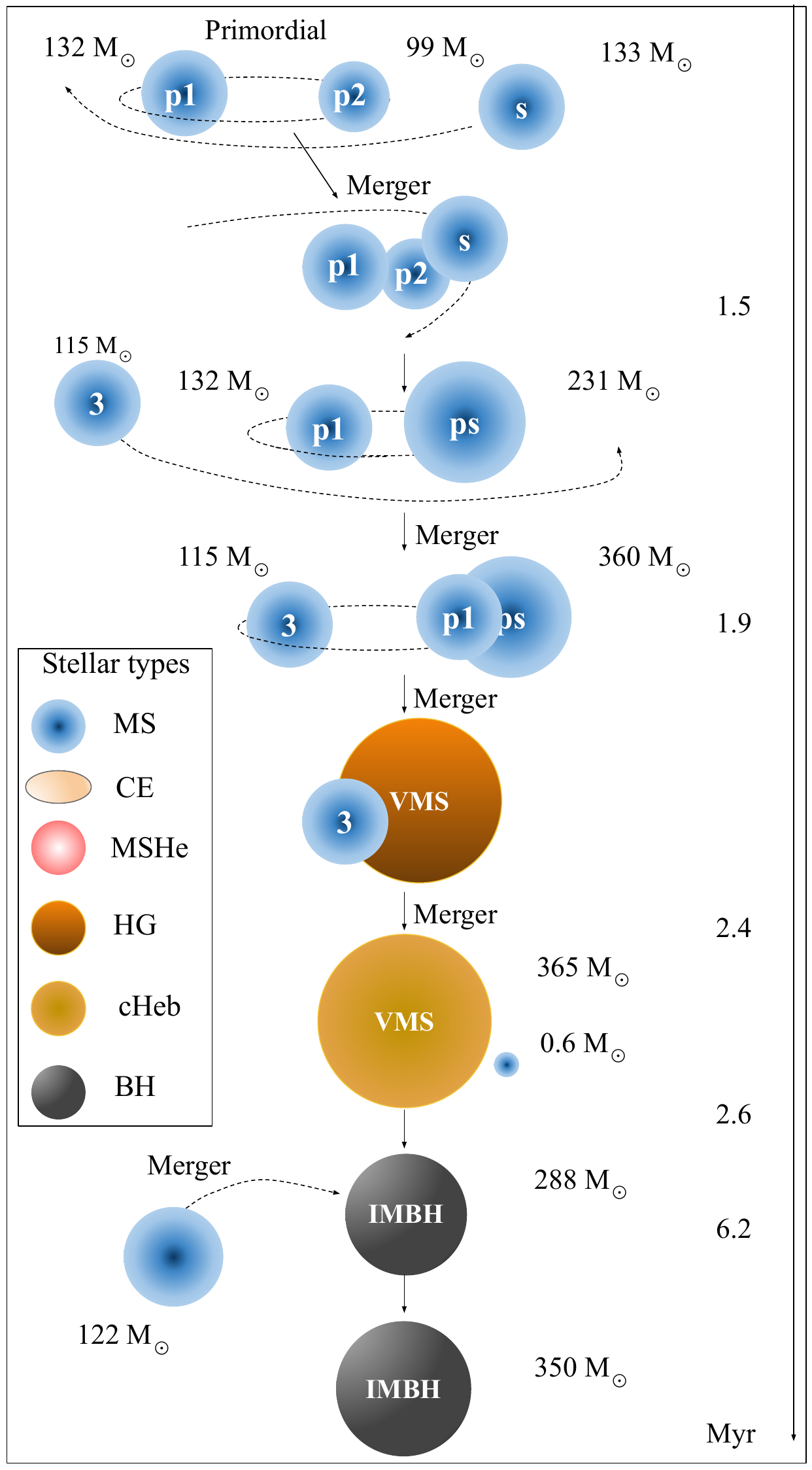}
    \caption{Formation of an IMBH in simulation with $N=120$k, $R_{\ham}=0.47$ pc, and $f_b=0.2$, realization ID 0. A sequence of massive star mergers triggers the formation of a very massive star (VMS) with mass $m_{\rm VMS}=365\Ms$ that directly collapse to an IMBH. The IMBH later accrete matter from another massive main sequence star, reaching a final mass of $m_{\rm IMBH}=350\Ms$. Different colors correspond to different evolutionary stages: main sequence (MS), common envelope (CE), naked main sequence He star (MSHe), Hertzsprung gap (HG), core He burning (cHeb), and black hole (BH).}
    \label{fig:fig3a}
\end{figure}

In the first example, an IMBH with final mass $m_{\rm IMBH} = 350\Ms$ forms in a cluster with $N=120$k stars, half-mass radius $R_\ham = 0.47$pc, and binary fraction $f_b=0.2$. The IMBH formation sequence is sketched in Figure \ref{fig:fig3a}. 

Initially, a primordial binary with component masses $m_{p1,p2} = (132 + 99)\Ms$ undergoes a series of strong interactions with a single MS star with mass $m_{s} = 133\Ms$ within the inner Myr of cluster evolution. The triple formed this way undergoes both phases of resonant interactions, with an exchange among the binary secondary and the third star, and a phase of hierarchical evolution, until the third body and the companion merge, leaving behind a new binary with component masses $m_{p1,ps} = (132+231)\Ms$, eccentricity $e \sim 0.001$ and semimajor axis $a \simeq 225$ R$_\odot$. After $1.8$ Myr, the binary captures a massive companion with mass $m_3 = 115\Ms$ that induces the collision of the two massive stars, eventually leaving behind a VMS with mass $m_{\rm VMS} = 360\Ms$, which forms a binary with $m_3$. The two binary components merge during the Hertzsprung-gap (HG) phase of the primary, leading to the formation of a VMS with total mass $m_{\rm VMS} = 365\Ms$. After capturing via a hyperbolic collision a small MS star ($\sim 0.7\Ms$) during the CHe burning phase, the VMS collapses to a BH with final mass $m_{\rm IMBH,1} = 288\Ms$ over a total time of $T_{\rm sim} = 2.5$ Myr. Within the subsequent $4$ Myr, the newborn IMBH collides with another massive MS star with mass $m_{\rm MS} = 122\Ms$, accreting a fraction $f_c = 0.5$ of its total mass and reaching a final IMBH mass of $m_{\rm IMBH} \simeq 350\Ms$. This case represents a clear example of how different formation channels, in this case stellar and star-BH mergers, concur to the IMBH seeding and growth.

\begin{figure}
    \centering
    \includegraphics[width=\columnwidth]{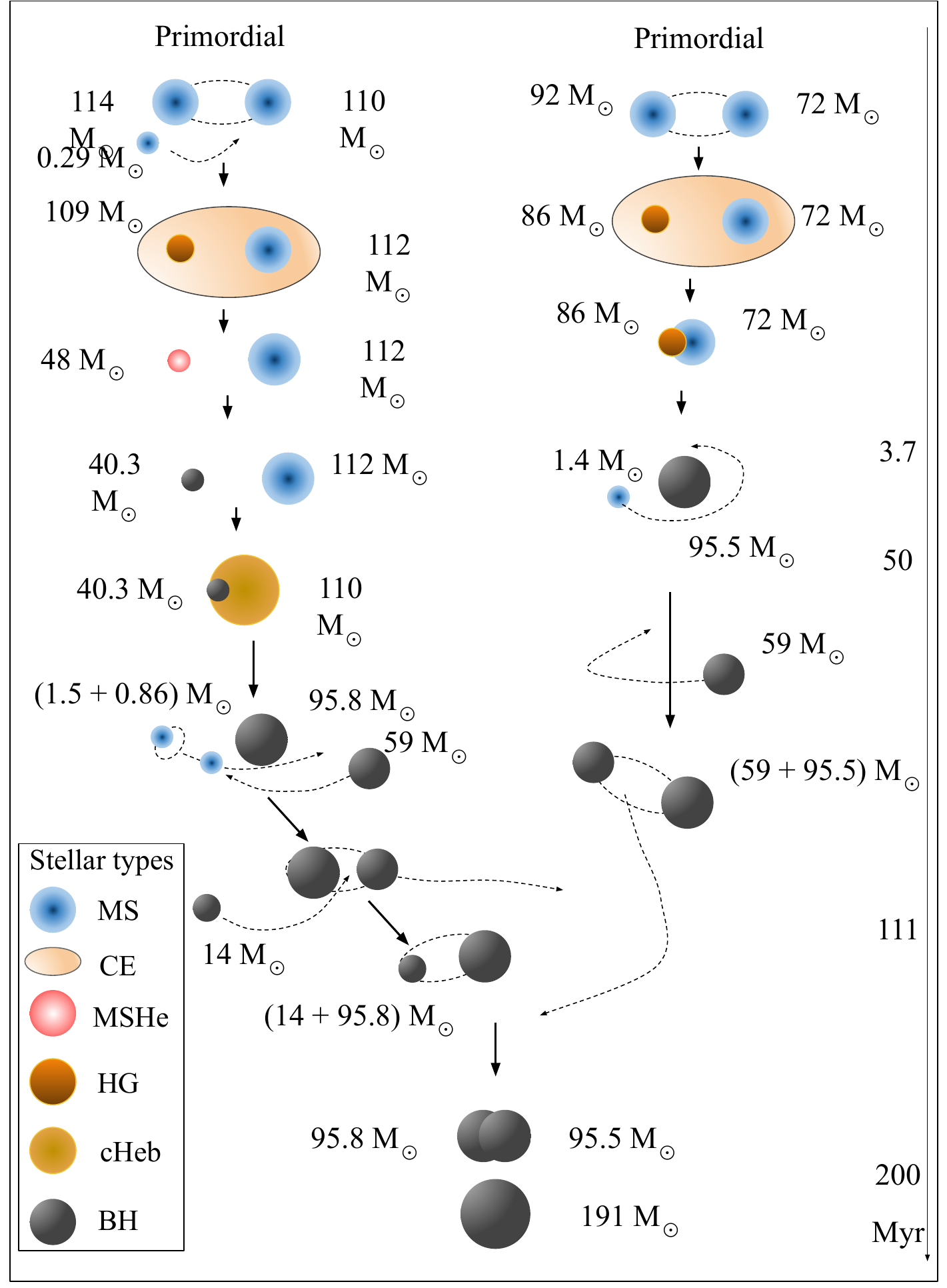}
    \caption{Formation of an IMBH in simulation with $N=120$k, $R_{\ham}=1.75$ pc, and $f_b=0.2$, realization ID 0. Two massive primordial binaries undergo common envelope that eventually lead to the formation of two nearly equal mass BHs ($m_{\rm BH}\sim 95\Ms$) that eventually find each other via a complex series of binary-binary interactions. The binary eventually merge and builds-up an IMBH with mass $m_{\rm IMBH} \simeq 191\Ms$. The color-coded legend is the same as in Figure \ref{fig:fig3a}.}
    \label{fig:fig3c}
\end{figure}

In the second example, instead, an IMBH with mass $m_{\rm IMBH} = 191\Ms$ form from the coalescence of two nearly equal mass BHs. As sketched in Figure \ref{fig:fig3c}, the two BHs with masses $\sim 95\Ms$ form from the evolution of two initially indipendent primordial binaries. After formation, the two BHs are part of different binaries and undergo many binary-single and binary-binary interactions before finding each other and merge after a time of $\sim 10^2$ Myr.

\subsubsection{Stellar mergers}
\begin{figure}
    \centering
    \includegraphics[width=\columnwidth]{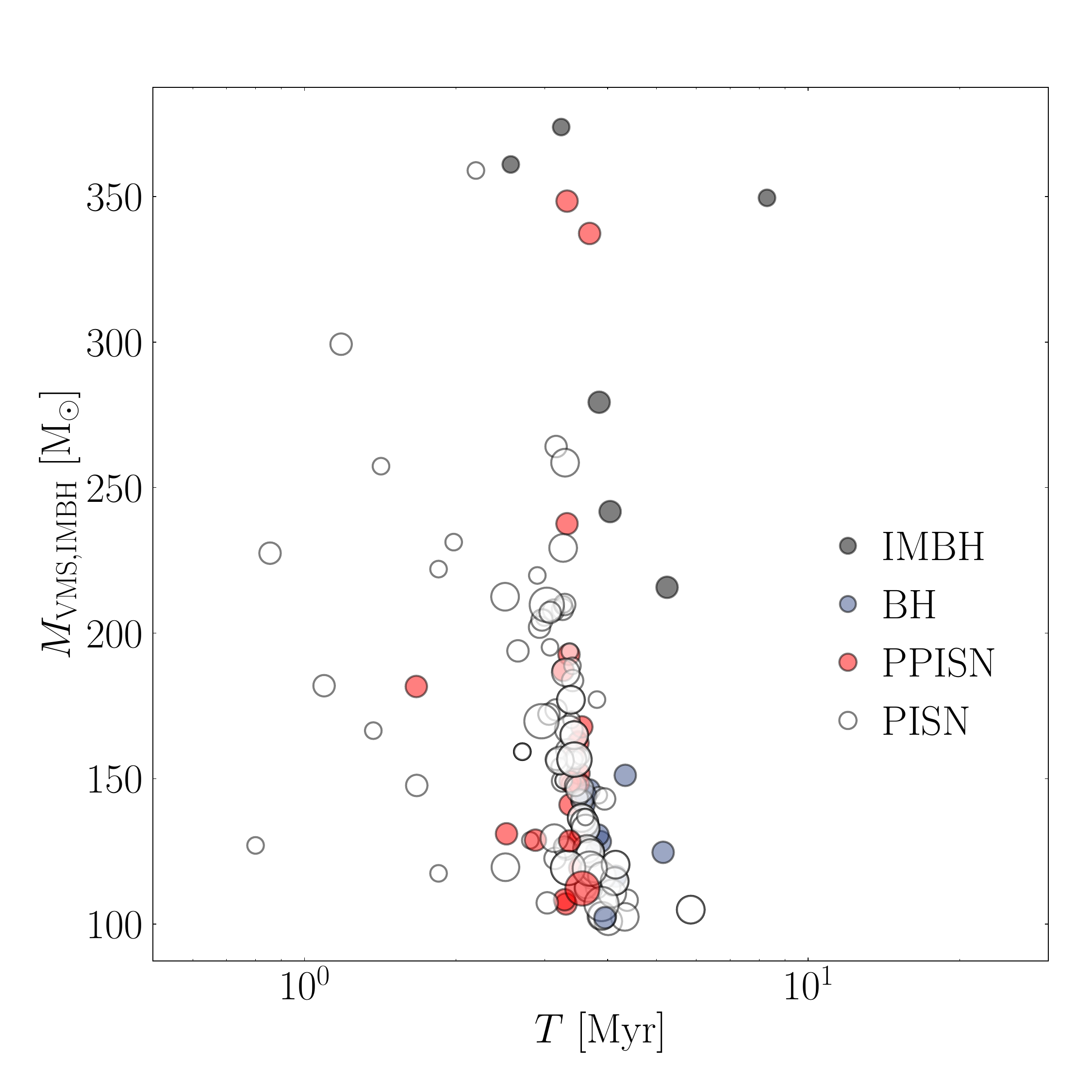}
    \caption{Formation time and final mass of VMSs formed in our models. We distinguish between VMS turning into an IMBH via direct collapse (black points), forming a "normal" BH (blue points), exploding via PPISN (red points), or via PISN, the latter leaving no remnant (white points).}
    \label{fig:VMS}
\end{figure}

In \dragonii models we find in total $104$ stellar mergers with a merger remnant mass heavier than $m_{\rm VMS}>90\Ms$, with $75\%$ of them involving primordial binaries. The typical mass of the merger product is a star with mass in the range $m_{\rm VMS} = 100-350\Ms$. In some cases, the same star undergoes 3-4 merging events with stars in different evolutionary phases. Figure \ref{fig:VMS} shows the post-merger mass as a function of the time at which the merger occurs for all simulations. The plot shows exclusively star-star coalescences, thus it excludes both star-BH and BH-BH merging events. Around $48\%$ of stellar mergers produce a massive MS star, $32\%$ produce a star in the HG, and a core-He burning star in the remaining $22\%$ of cases. 
The formation of a VMS ($m_{\rm VMS}> 150\Ms$) eventually leaves to either no remnant owing to PISN ($\sim 23$ cases), a remnant with mass $m_{\rm BH} = 40.5\Ms$ owing to PPISN ($\sim 64$ cases), or an IMBH ($\sim 2$ cases). 

Comparing models with same $R_\ham$ and different binary fraction, we find that models with $f_b=0.2$ host a number of mergers 2-5 times larger than the case $f_b=0.05$, a reflection of the fact that most of the mergers involve primordial binaries. 

Noteworthy, the two IMBHs form in the densest simulated clusters, i.e. those with $R_\ham = 0.47$ pc and $N=(1.2-3)\times 10^3$, which are also those with the shortest mass-segregation time ($T_{\rm seg} \sim 0.3-0.4$ Myr), much shorter than the typical BH formation time ($>2$ Myr).

\subsubsection{Star-black hole collisions}

Among all simulations, we find 454 star-BH merger events, the vast majority of which ($72\%$) lead to the formation of BHs with a final mass $m_{\rm BH}<40.5\Ms$, thus they will remain mixed with the population of "ordinary" BHs that never experienced stellar accretion episodes. The remaining mergers leave behind, instead, BHs with a mass falling in the upper-mass gap. More in detail, around $18\%$ of these events trigger the formation of a final BH with a mass in the range $40.5 < m_{\rm BH}/\Ms < 60$, $6\%$ form BHs with masses in the $60 < m_{\rm BH}/\Ms < 70$ mass range, and the remaining $\sim 4\%$ produces BHs heavier than $m_{\rm BH} > 70\Ms$. Stars involved in a star-BH merger are in different evolutionary stages: HG ($40.1\%$), core He burning ($45.2\%$), MS ($5.5\%$), early/late asymptotic giant branch (AGB, $9\%$), giant branch (GB, $1.1\%$), and HG naked He star ($0.2\%)$.

Note that we have two different type of star-BH accretion events: one purely dynamical and one induced by stellar evolution. In the purely dynamical case, we have two possibilities: either the BH captures a MS star in a orbit such that the star fills its Roche lobe, or the orbit is sufficiently tight and eccentric that the BH crashes onto the star. In any case, the BH accretes a fraction $f_c$ of the star mass. In the stellar evolution-driven case, instead, the star fills its Roche lobe, mainly when inflating during the HG or the core He burning phase. Even in such case, though, in \nbsix it is assumed that the BH eats up a fraction $f_c$ of the star mass. Therefore, the stellar type is likely the parameter that better identify the two types of star-BH accretion/merger events. 

\begin{figure*}
    \centering
    \includegraphics[width=0.45\textwidth]{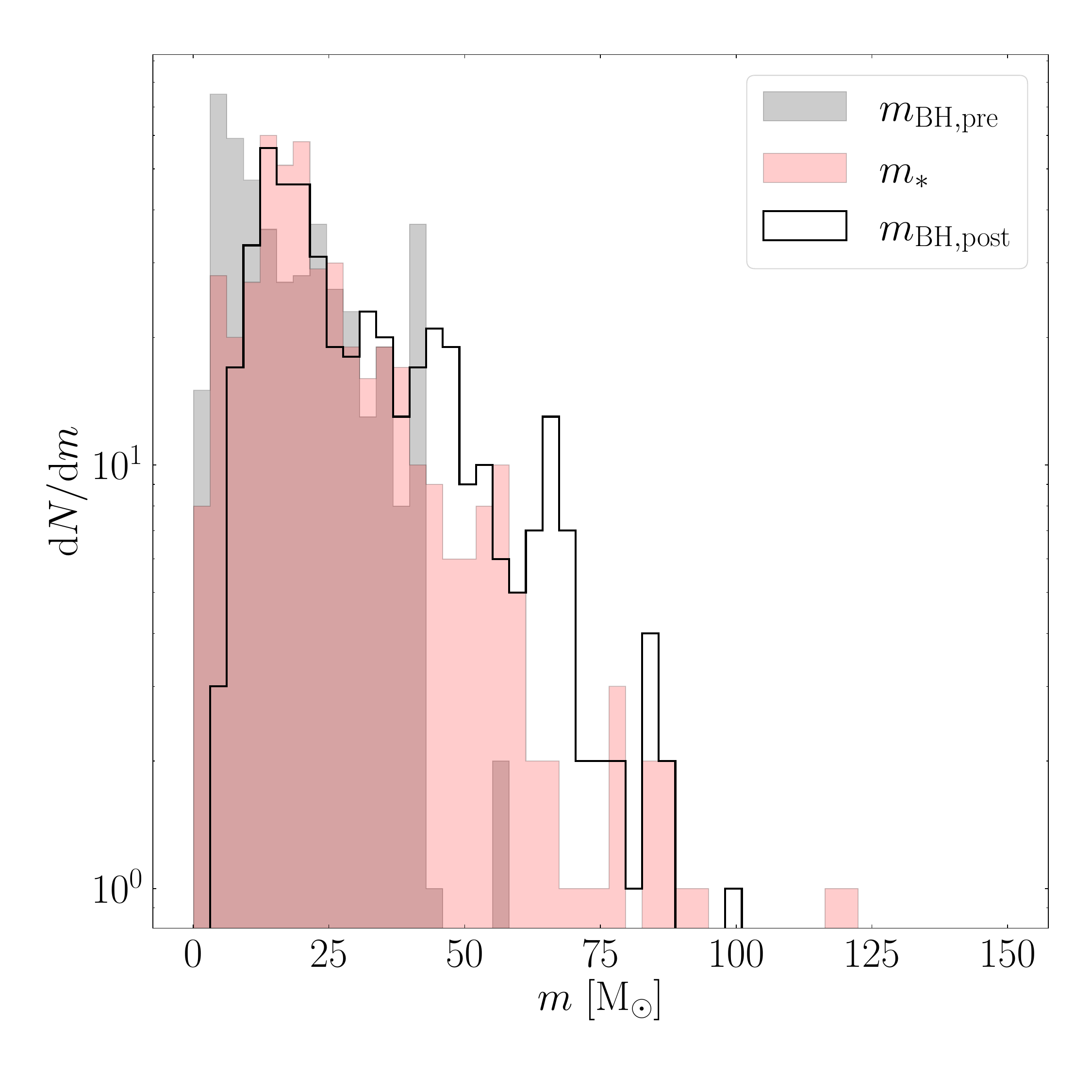}
    \includegraphics[width=0.45\textwidth]{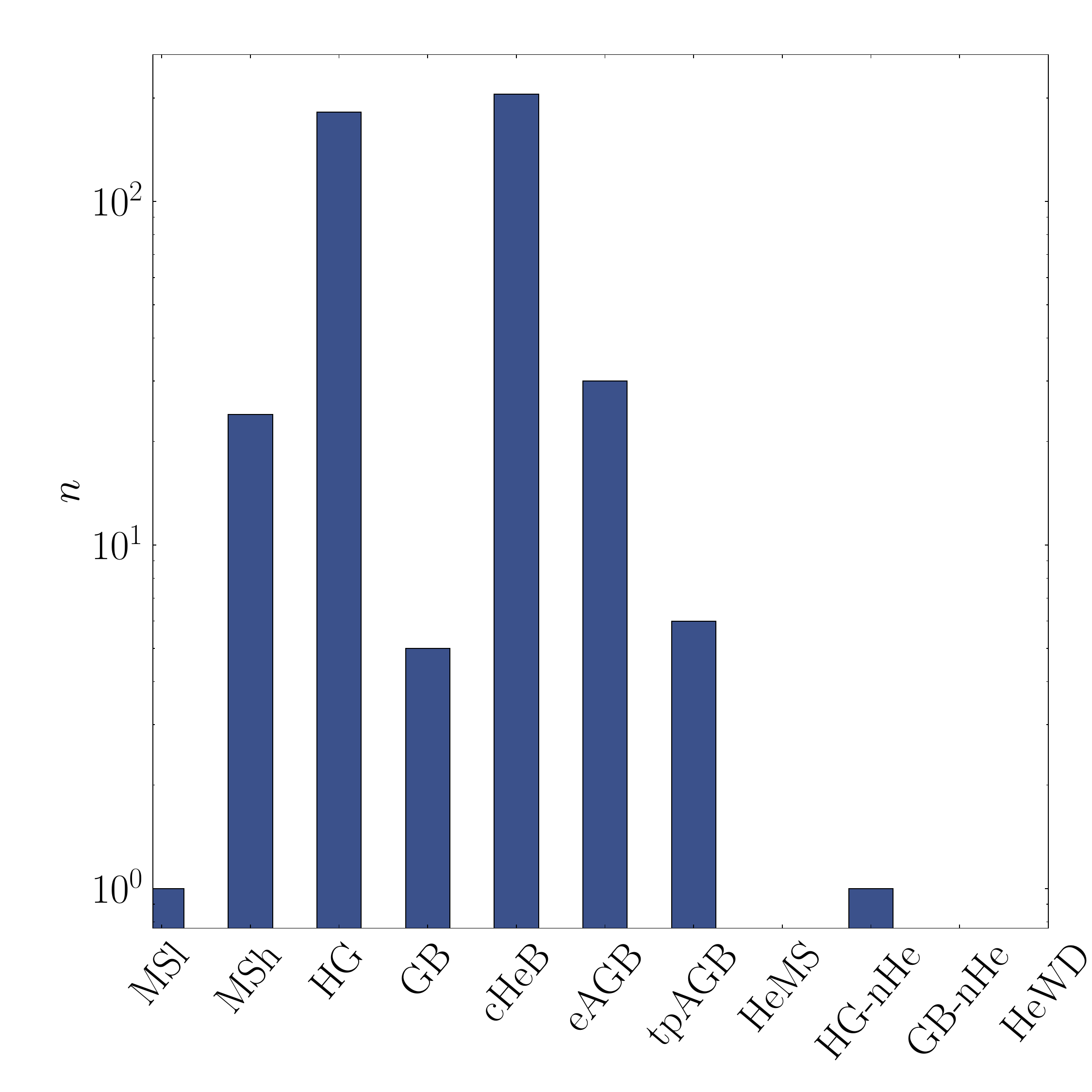}
    \caption{Left panel: mass distribution of BHs (grey filled steps) and stars (red filled steps) involved in a star-BH merger compared to the final BH mass distribution (black open steps). Right panel: stellar type of stars involved in a star-BH merger.}
    \label{fig:gap}
\end{figure*}

Figure \ref{fig:gap} shows the mass distribution of the merging star, and the BH before/after the merger, and the stellar type of the stars involved in the process. 

Two events contribute to IMBH seeding or growh, one of them involves a $m_{\rm BH}=40.5\Ms$ BH that accretes a core He burning star with mass $m_{\rm VMS} = 133\Ms$, previously formed via a complex sequence of stellar mergers triggered by binary-binary and binary-single interactions. In such case, the IMBH mass is $m_{\rm IMBH} = 107\Ms$. The second event, which we do not show in the histogram to ensure an optimal visibility, involves an IMBH with mass $m_{\rm IMBH} = 288\Ms$ and a MS star with mass $m_* \simeq 122\Ms$. None of all other interactions lead to the formation of an IMBH, partly owing to our choice to set the accretion factor to $f_c=0.5$. Adopting $f_c = 1$ would have lead to an additional population of $\sim 20$ IMBHs with a mass at formation in the range $m_{\rm IMBH} = 100-160\Ms$. 

\subsubsection{Black hole mergers}

The remaining 5 IMBHs in \dragonii clusters form via BH-BH mergers, all involving upper mass-gap BHs. This highlights the fundamental impact of star-BH accretion events, because they are the main channel through which mass-gap BHs form. Interestingly, all the BH mergers involved in the IMBH buildup have progenitor stars originally in a primordial binary, thus highlighting the crucial role of binary dynamics in the IMBH formation process.
At formation, these 5 IMBHs have masses in the range $m_{\rm IMBH}\simeq(140-232)\Ms$ and, in case of negligible GW recoil, further increase up to $m_{\rm IMBH}\simeq(160-260)\Ms$ via one or two repeated (hierarchical) merger events, after being dynamically ejected from the cluster. In the case of zero GW recoil, among all IMBHs in \dragonii models, only one is ejected from the cluster as a single object. All the other are ejected with a companion and undergo merger within a Hubble time. In two cases, the IMBH undergoes two/three mergers inside the cluster and forms a binary with another BH that is eventually ejected from the cluster, merging in the field within a Hubble time.

\subsubsection{The link between formation channels, formation times, and the intermediate-mass black hole mass}

Despite our sample is rather small, the fact that \dragonii IMBHs form via all the proposed formation channels can help to provide a possible answer to the intriguing question
\newline
"Is there a link between the IMBH seeding process and the environment in which this happens?"
\newline

\begin{figure}
    \centering
    \includegraphics[width=\columnwidth]{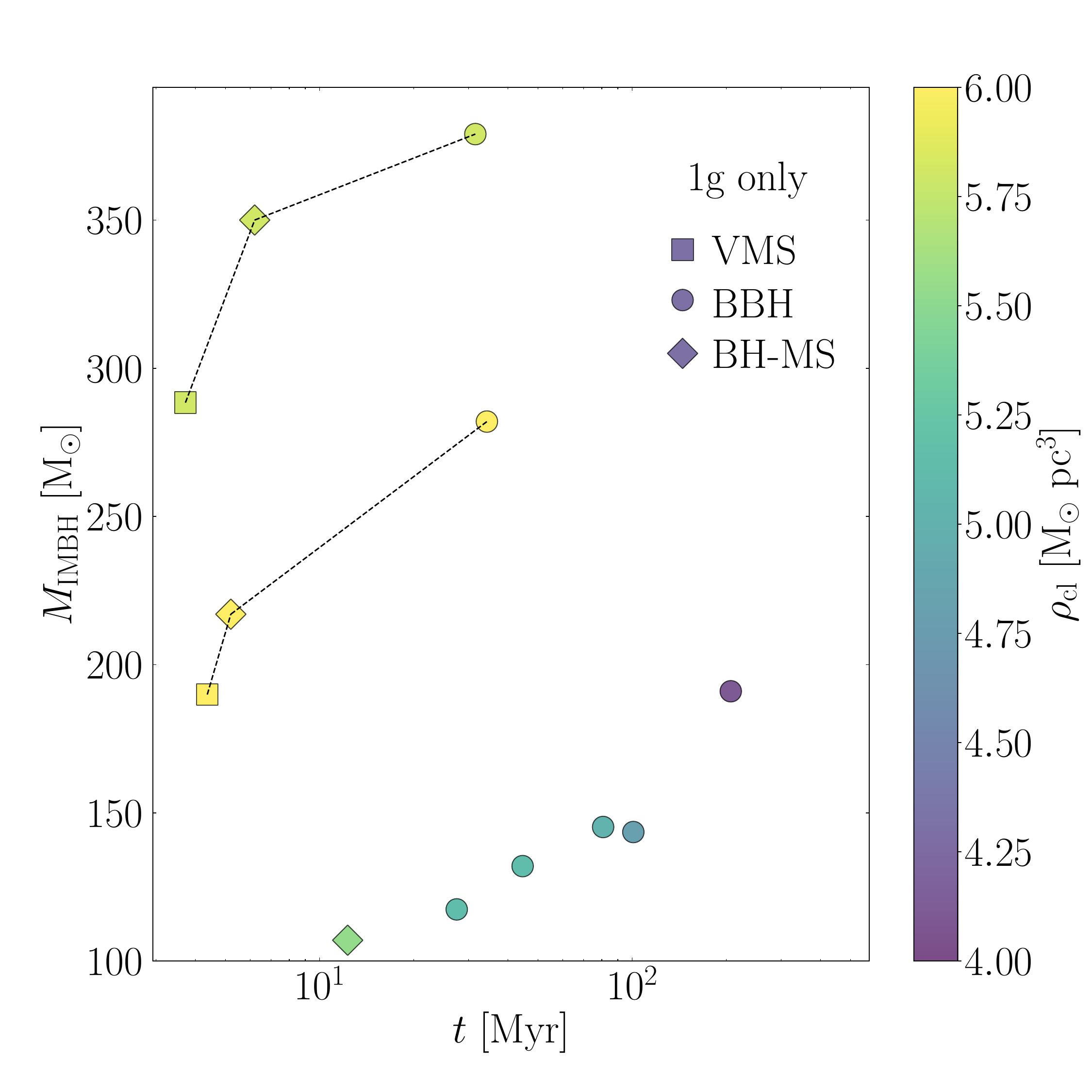}\\
    \caption{Top panel: Time and mass of IMBHs at formation, i.e. when the IMBH mass exceeds $10^2\Ms$ for the first time. We distinguish between IMBHs forming via stellar collisions (VMS, squares), binary BH mergers (BH-BH, circles), and star-BH accretion (BH-MS, diamonds). In two cases, the IMBH accretes stellar material from a massive star after seeding. The consequent mass increase is highlighted with dashed lines. The color coding identifies the initial density of the host cluster. Only 1g-IMBHs are considered here.}
    \label{fig:imbF}
\end{figure}

Figure \ref{fig:imbF} shows the IMBH mass as a function of time for different formation channels from the first time the IMBH mass exceeds $10^2\Ms$ and until the first BH merger event develops. In other words, we exclude from the plot IMBHs older than the second generation (2g), because GW recoil drastically reduce the probability for multiple generation mergers, as discussed in Section \ref{sec:reten}.  

From the plot, it seems that there is a striking relation between the structure of the host cluster and the IMBH formation process. The densest clusters ($\rho_{\rm cl} > 3\times 10^5\Ms~{\rm pc}^{-3}$) favour the formation of IMBHs via stellar collisions on the short timescales ($<10$ Myr) and nurture the most massive IMBHs in our sample. IMBHs in these clusters further grow via accretion of stellar material and coalescence with stellar BHs on timescales $<100$ Myr \citep[see also][]{2022MNRAS.514.5879M}. 
In lower density clusters, instead, IMBHs form on longer timescales ($10-300$ Myr) via star-BH accretion and BBH mergers. In such case, Figure \ref{fig:imbF} clearly shows a trend, namely that the looser the cluster the longer the formation time and the heavier the IMBH seed mass. 

This difference may be related to the core-collapse process, a mechanism driven by mass-segregation and relaxation according to which the cluster core contracts and its density increases up to a maximum point, i.e. the core-collapse. The time at which core-collapse occurs is generally a fraction of the relaxation time, $t_{\rm cc} = 0.2 T_{\rm rlx}$ \citep{2002ApJ...576..899P,2014MNRAS.439.1003F}. We find that in clusters with an initial density $>3\times 10^5\Ms~{\rm pc}$ the core-collapse occurs before stellar BH forms or massive stars undergo PISN and PPISN, i.e. $t_{\rm BH} \sim 4$ Myr. This supports the idea that core-collapse facilitate the collision of the stars before they collapse to BH or undergo PISN. 

In the case of clusters less dense that $3\times 10^5\Ms~{\rm pc}$, we also note that the smaller the density the larger the IMBH mass. This may be due to the fact that in low-density clusters, where interactions are less energetic and less frequent, the ejection of the most massive BHs via the so-called BH-burning process \citep[e.g.][]{2013MNRAS.432.2779B,2018MNRAS.479.4652A,2020IAUS..351..357K} is less effective. As a consequence, the heaviest BHs in the loosest clusters in our sample have more time to hang around in the cluster and pair-up, as in the case of model IBH\_Rh1.75f20N120k. 

\section{Discussion}
\label{sec:disc}

\subsection{Newtonian versus relativistic dynamics: intermediate-mass black hole retention and hierarchical mergers frequency}
\label{sec:reten}

In this work, we want to assess the competing role of Newtonian and relativistic dynamics in determining BH retention and IMBH seeding and growth, thus we adopt the following multi-stepped procedure: a) run all cluster simulations assuming zero GW recoil to verify the possible development of multiple mergers and quantify the impact of Newtonian dynamics on the retention of BH merger remnants, b) quantify the retention probability of remnant BHs, c) re-run models in which BHs undergo repeated mergers with GW recoil enabled.

\subsubsection{Newtonian dynamics}

Regardless of the formation scenario, an IMBH seed that upon formation is retained in its parent cluster will likely undergo mass-segregation and quickly settles in the cluster centre possibly capturing a companion \citep[e.g.][]{2016ApJ...819...70M,2019arXiv190500902A}. The newly formed binary will undergo frequent interactions with surrounding cluster members with mass $m_p$, at a rate
\begin{equation}
    \dot{n}_{2-1} \sim n \sigma \pi a^2(1-e)^2 \left(1+\frac{2G(m_1+m_2+m_p)}{a(1-e)\sigma^2}\right), 
\end{equation}
where $n$ is the cluster number density, $\sigma$ the velocity dispersion, $m_{1,2}$ the mass of binary components, and $a$ the binary semimajor axis. If the binary is hard, i.e. $a \ll 2G(m_1+m_2)/\sigma^2$, or highly eccentric, the timescale for these interaction is roughly given by
\begin{align}
    t_{2-1} &\sim 6{\rm Myr} \left(\frac{n}{10^5{\rm pc}^{-3}}\right)^{-1}\left(\frac{\sigma}{20{\rm km~s^{-1}}}\right) \times \\\nonumber
    &\times \left(\frac{m_1+m_2+m_p}{240\Ms}\right)^{-1} \left(\frac{a}{1{\rm AU}}\right)^{-1} (1-e),
\end{align}
therefore much shorter than the typical cluster lifetime. Repeated binary-single interactions can have an important effect on the binary evolution: on the one hand, they can extract orbital energy and harden the binary \citep[\textit{Heggie's law},][]{1975MNRAS.173..729H}, but, on the other hand, they can become so violent to eject the binary from the cluster, halting the IMBH growth \citep{2022MNRAS.514.5879M}. 

The typical escape velocity of clusters described by a \cite{1966AJ.....71...64K} model can be conveniently expressed as \citep[see e.g.][]{2022ApJ...927..231F}
\begin{equation}
    v_{\rm esc} = 2\sqrt{\log(1/c)/\pi}\left(1-c\right)^{-1/2} \left(\frac{GM}{R_\ham}\right)^{1/2}, 
\end{equation}
where $c = R_c/R_\ham$ is the ratio between the core and half-mass radius of the cluster. In \dragonii models, we find that such parameter attains values $c=0.2\pm 0.1$ within the whole simulation time and regardless of the initial conditions. Therefore, the escape velocity can be rewritten as 
\begin{equation}
    v_{\rm esc} = (34\pm 3){\rm km/s} \left(\frac{M}{10^5\Ms}\right)^{1/2}\left(\frac{R_\ham}{1{\rm pc}}\right)^{-1/2}.
    \label{eq:vesc}
\end{equation}
In all \dragonii clusters the escape velocity remains below $v_{\rm esc} < 50$ km$/$s, with the loosest and smallest clusters attaining values in the $8-20$ km/s range. This relatively small escape velocity has huge implications on the IMBH evolution. In fact, even when GW recoil is not taken into account, all \dragonii IMBHs are ejected from the parent cluster after a violent interaction with a perturber. 

A clear example is a simulation with $N = 300$ k, $R_\ham = 0.47$ pc, and $f_b = 0.2$, in which a binary with mass $m_1 + m_2 =
(240 + 38)\Ms$ undergoes a strong scattering with a BH with mass
$m_p = 44 \Ms$, which reduces the binary semimajor axis from $a = 0.35$ AU to $a_{\rm fin} = 0.24$ AU and impart to the binary a recoil with amplitude $v_{\rm rec} = 85$ km s$^{-1}$. From a theoretical standpoint, a binary undergoing a close interaction with a perturber with mass $m_p$ and consequently shrinking from $a$ to $a_{\rm fin}$ receives a kick
\citep{1975MNRAS.173..729H,1993ApJS...85..347H,1993ApJ...415..631S,1993ApJ...403..271G,1996NewA....1...35Q, 2016ApJ...831..187A, 2022MNRAS.514.5879M} 
\begin{align}
    v_{\rm rec} &= \left[ \frac{Gm_1m_2}{a_{\rm fin}(m_1+m_2)}\frac{m_p}{m_1+m_2+m_p}\left(1-\frac{a_{\rm fin}}{a}\right) \right]^{1/2}= \nonumber \\
    &= 37.1\kms \left(\frac{\mu}{26\Ms}\right)^{1/2}\left(\frac{q_p}{0.12}\right)^{1/2}\left(\frac{a_{\rm fin}}{1{\rm AU}}\right)^{-1/2}\left(1 - \frac{x_{\rm fin}}{0.5}\right)^{1/2},
    \label{eq:8}
\end{align}
where $\mu = m_1m_2/(m_1+m_2)$ and $q_p = m_p/(m_1+m_2+m_p)$. This equation returns a value $v_{\rm rec} \simeq 72$ km s$^{-1}$ for the aforementioned example.
This implies that as long as at least one heavy ($m_p > 10\Ms$) perturber remains in the cluster, Newtonian dynamics, in particular close binary-single scatterings, can represent a serious threat to the IMBH retention. 

Our analysis highlights the extreme importance of Newtonian dynamics in determining the evacuation of BHs from the parent cluster. 

\subsubsection{The impact of black hole natal spins and relativistic recoil on the properties of intermediate-mass black holes}
\label{sec:spin}

In order to determine the possible properties of IMBHs and their retention probability in \dragonii models, we implement the following simple model to take into account the impact of spins: 
\begin{itemize}
    \item If a stellar BH involved in the IMBH build-up formed from a single star or from a ``non-interacting'' binary, we assign a spin of $\chi_{\rm BH} = 0.01$ \citep{2002A&A...381..923S,2019ApJ...881L...1F}. 
        
    \item In the two cases in which an IMBH forms from the collapse of a VMS assembled via stellar mergers, we assign an initial spin of 0.5. The choice is motivated by the fact that the particularly complex formation processes that lead to the IMBH formation make the IMBH natal spin practically unpredictable. We note that this choice has no effect on our results though, because both IMBHs accretes material from a stellar companion and we assume that this spin-up the IMBH as detailed in the following point.

    \item If the IMBH feeds on a stellar companion, or if its progenitors are upper-mass gap BHs, i.e. they underwent mass accretion at some point, we assign a spin drawn from a flat distribution in the range $\chi_{\rm BH} = 0.8-1$ \citep[see e.g.][]{2018A&A...616A..28Q,2020A&A...635A..97B,2020A&A...636A.104B,2020ApJ...892...13S}. 

    \item If the IMBH progenitor is a BH formed in a primordial binary, we assign a small spin ($\chi_{\rm BH} = 0.01$) if it is the firstborn or a spin in the range $\chi_{\rm BH} = 0.1-1$ \citep{2018A&A...616A..28Q, 2020A&A...635A..97B} otherwise. 

    \item If the IMBH formed from a BBH merger, the IMBH spin and mass are calculated according to \cite{2017PhRvD..95f4024J} fitting formulae \citep[see also]{2020ApJ...894..133A}.
\end{itemize}
Note that this model is applied in post-process to the simulation data.

To keep track of the IMBH-BH merging history, we label an IMBH as first generation (1g) if it did not undergo any merger with another compact object. IMBHs formed out of VMS collapse or star-BH accretion are considered 1g. Second generation (2g) and higher generation IMBHs are those that underwent multiple mergers with other compact objects. In \dragonii models, all merging companions are stellar BHs. 

\begin{figure*}
    \centering
    \includegraphics[width=1.7\columnwidth]{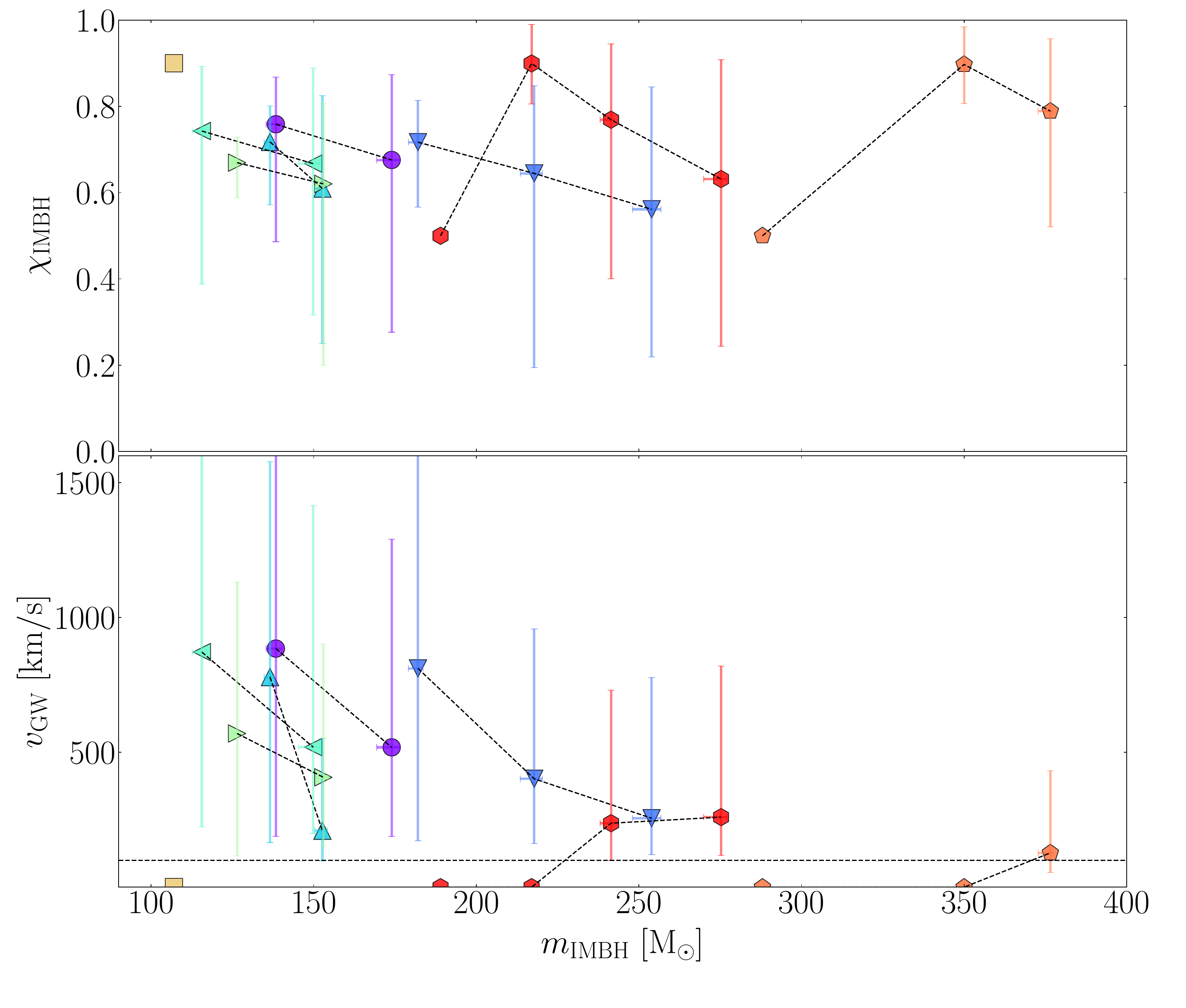}
    \caption{Evolutionary tracks of IMBHs in \dragonii models in terms of IMBH mass (x-axis), spin (top panel), and recoil kick (bottom panel). IMBHs that form via stellar mergers or BH-star interactions are assumed to have zero kicks at formation. The error bars enclose the $95$th percentile. Dashed lines identify the evolutionary path of each IMBH, whilst the horizontal dashed line in the bottom panel marks a velocity threshold of $100$ km$/$s, i.e. typical value of galactic nuclei similar to the Milky Way.}
    \label{fig:IBHs}
\end{figure*}

Figure \ref{fig:IBHs} shows the masses and spins of \dragonii IMBHs assuming zero GW recoil. It appears evident that, upon our assumptions, IMBHs in \dragonii clusters generally form with a high spin ($\chi_{\rm IMBH} > 0.6$), unless they form from the collapse of a VMS. Even in such a case, the accretion of matter, which likely spins-up the IMBH, occurs on a sufficiently short timescale ($t\lesssim 8$ Myr) to make rather unlikely their observation as low-spin objects.
In the case of IMBHs forming via multiple BH mergers, note that the IMBH spin decreases at increasing the merger generation \citep[see also][]{2021A&A...652A..54A}. 

Table \ref{tab:tabimbh} summarizes the main properties of \dragonii IMBHs in terms of generation, masses, spins, and recoil velocity at $95\%$ confidence level. These quantities are calculated drawing for each merging event 10,000 times the spin amplitude of the merging components and assuming for the spin directions an isotropic distribution.
Looking at the Table, we see that GW recoil has no effect on the IMBH formation probability, because all IMBHs in \dragonii clusters form either via stellar collapse or have a 1g BH progenitor. Nonetheless, GW recoil crucially affects second and higher generation IMBHs, which typically receive a kick, $v_{\rm GW} = (200 - 800) \kms$, much larger than the escape velocity from the parent cluster, typically $v_{\rm esc} < 50\kms$. 
Therefore, the inclusion of GW recoil affects 7 out of 8 IMBHs in our simulations, avoiding both: a) the formation of IMBH-BH binaries that merge after dynamical ejection, a process involving 5 IMBHs in our sample, and b) the development of multiple BH mergers inside the cluster (2 IMBHs). 
The remaining IMBH is ejected from the cluster as a single object after a strong resonant interaction with other two, fairly massive ($>30\Ms$), BHs. 
As a consequence, we find that the number of merging events involving an IMBH decreases from 9 in the no-recoil case, to just 2, despite this represents the lowest value possible. The possible detection of GWs emitted from IMBH-BH binaries with future detectors, especially those operating in the deci-Hz frequency band,  could help shed a light on the IMBH formation efficiency and retention probability \citep[see e.g.][]{2011GReGr..43..485G,2020CQGra..37u5011A,2021A&A...652A..54A,2022A&A...659A..84A}.

\begin{table*}
    \centering
    \renewcommand{\arraystretch}{1.5}
    \begin{tabular}{ccccccc}
    \hline
    \hline
    $R_\ham$ & $f_b$ & $N$ & $M_{\rm IMBH}$ & $\chi_{\rm IMBH}$ & $v_{\rm GW}$ &  $\#$ of \vspace{-0.15cm} \\
    pc       &   $10^{-2}$  & $10^5$& M$_\odot$    & & km$/$s & gen\\
    \hline
       1.75&  5 & 10  &  $138.4^{+1.8}_{-3.0}$ & $0.759^{+0.109}_{-0.272}$ & $885^{+1059}_{-697 }$ & 2g \\
       1.75&  5 & 10  &  $174.0^{+2.6}_{-4.6}$ & $0.676^{+0.198}_{-0.399}$ & $519^{+773 }_{-330 }$ & 3g \\
       1.75& 20 &  1.2  &  $181.8^{+1.8}_{-2.7}$ & $0.717^{+0.097}_{-0.151}$ & $811^{+820 }_{-640 }$ & 2g \\
       1.75& 20 &  1.2  &  $217.8^{+2.5}_{-4.3}$ & $0.645^{+0.203}_{-0.450}$ & $402^{+557 }_{-240 }$ & 3g \\
       1.75& 20 &  1.2  &  $253.9^{+2.9}_{-5.9}$ & $0.561^{+0.284}_{-0.342}$ & $256^{+522 }_{-135 }$ & 4g \\
       1.75& 20 &  6  &  $136.6^{+1.2}_{-1.9}$ & $0.718^{+0.083}_{-0.146}$ & $779^{+799 }_{-614 }$ & 2g \\
       1.75& 20 &  6  &  $152.7^{+1.5}_{-2.4}$ & $0.610^{+0.215}_{-0.360}$ & $210^{+341 }_{-112 }$ & 3g \\
       0.80& 20 &  1.2  &  $115.6^{+1.3}_{-3.0}$ & $0.743^{+0.150}_{-0.355}$ & $872^{+1069}_{-649 }$ & 2g \\
       0.80& 20 &  1.2  &  $149.8^{+2.0}_{-4.6}$ & $0.667^{+0.221}_{-0.351}$ & $519^{+896 }_{-320 }$ & 3g \\
       0.80& 20 &  1.2  &  $126.4^{+0.7}_{-1.0}$ & $0.670^{+0.059}_{-0.081}$ & $570^{+561 }_{-454 }$ & 2g \\
       0.80& 20 &  1.2  &  $153.0^{+1.4}_{-2.1}$ & $0.620^{+0.187}_{-0.420}$ & $407^{+495 }_{-254 }$ & 3g \\
       0.80& 20 &  3.  &  $107$                 & $0.01$                    & $-$                   & 1g \\
       0.47& 20 &  1.2  &  $288$                 & $0.5$                     & $-$                   & 1g \\
       0.47& 20 &  1.2  &  $350$                 & $0.8-1$                   & $-$                   & 1g* \\
       0.47& 20 &  1.2  &  $376.5^{+0.8}_{-3.7}$ & $0.789^{+0.168}_{-0.269}$ & $127^{+305 }_{- 72 }$ & 2g \\
       0.47& 20 &  3  &  $189$                 & $0.5$                     & $-$                   & 1g \\
       0.47& 20 &  3  &  $217$                 & $0.8-1$                   & $-$                   & 1g*  \\
       0.47& 20 &  3  &  $241.4^{+0.8}_{-3.3}$ & $0.769^{+0.176}_{-0.369}$ & $237^{+494 }_{-138 }$ & 2g \\
       0.47& 20 &  3  &  $275.3^{+1.8}_{-5.4}$ & $0.632^{+0.276}_{-0.388}$ & $260^{+561 }_{-143 }$ & 3g \vspace{0.1cm}\\
    \hline
    \end{tabular}
    \caption{Properties of the IMBH in \dragonii models. Col. 1-3: half-mass radius, binary fraction, and number of stars. Col. 4-6: median values of the mass, spin, and GW recoil of IMBHs. Errors represent the 95th percentile. Col. 7: BH generation. The asterisk marks a 1st generation IMBH -- formed via stellar mergers -- that grew through a merger with a massive star.}
    \label{tab:tabimbh}
\end{table*}

\subsubsection{Simulations implementing a self-consistent treatment for gravitational recoil}
\label{sec:self}

The post-process treatment applied to simulation data provides an effective way to place constraints on the IMBH retention probability without the need to explore the wide associated parameter space. Nonetheless, a fully self-consistent simulation implementing also GW recoils would provide useful insights on, e.g. the impact of the IMBH displacement onto the development of new merging events. 

To prove the impact of GW recoil in a self-consistent way, we focus on the two models in which the IMBH undergoes repeated mergers, namely models IBH\_Rh1.75f20N120k, which ultimately form a 4g-IMBH, and IBH\_Rh0.47f20N300k, which instead leads to a 3g-IMBH.  

Practically speaking, we restart the simulation from the snapshot immediately before the merging event and apply to the merger remnant a kick. For simplicity, rather than extracting the kick from a distribution we assign the merger a given kick, as described below. Generally, we adopt a GW kick sufficiently small to ensure the IMBH retention after the merger. This choice permits us to investigate whether the IMBH can be retained in the cluster, it further grows, or it is anyway ejected owing to Newtonian or relativistic effects.

\paragraph*{Model ID: IBH\_Rh1.75f20N120k}
The IMBH in this model forms from the merger of two upper mass-gap BHs with masses $m_{\rm BH1}+m_{\rm BH2} = (95.5+95.8)\Ms$. Therefore, the IMBH is already 2g at formation, and receives a kick $v_{\rm rec} > 171\kms$ at $95\%$ confidence level (see Table \ref{tab:tabimbh}). For comparison, the cluster escape velocity at the time of the merger is around $v_{\rm esc} = 12\kms$. 

Adopting the spin model described in Section \ref{sec:spin}, based on stellar evolution models, we find that the IMBH has a tiny fraction ($P_{20}<0.2\%$) to receive a kick $v_{\rm GW} < 20\kms$. However, if the IMBH progenitors have negligible spins for some reason, for example if the IMBH progenitor is slowly rotating and the angular momentum transport is essentially driven by meridional currents \citep{1992A&A...265..115Z, 2020A&A...636A.104B}, the probability for $v_{\rm GW}<20\kms$($5\kms$) rises up to $84\%$($21\%$), significantly increasing the IMBH retention probability. 

Therefore, we re-run the simulation and assign to the IMBH promptly after formation a GW kick of either $v_{\rm GW} = 5\kms$ (small kick) or $20\kms$ (large kick). As expected, in the large kick model, the kick of $v_{\rm GW} = 20\kms$ exceeds the cluster escape velocity and the IMBH promptly leaves the cluster.

In the small kick model, where $v_{\rm GW} = 5\kms$, the 2g-IMBH is retained in the cluster and sinks back to the cluster centre where, after a long series of interactions with other stellar BHs, captures a BH with mass $m_{\rm BH}=28\Ms$ and is ejected from the cluster with a velocity of just $15.3\kms$. The ejected IMBH-BH binary has an eccentricity $e=0.57$ and a period of $P=190$ days, and a corresponding merger time $t_{\rm GW} \sim 10^3$ Hubble times. 
For the sake of comparison, in the zero GW recoil model, the IMBH pairs with a BH with mass $m_{\rm BH} = 40.5\Ms$ and is ejected from the cluster, merging within a Hubble time (see Appendix \ref{sec:IMevo}). 

\paragraph*{Model ID: IBH\_Rh0.47f20N300k}
Let us now consider the other model, named IBH\_Rh0.47f20N300k. Since the IMBH in this model forms via stellar collisions, its mass at birth is fairly large $m_{\rm IMBH} = 217\Ms$. After only 17 Myr, when the cluster escape velocity is around $v_{\rm esc} = 46.5\kms$, this 1g-IMBH merges with an upper mass-gap BH with mass $m_{\rm BH} = 51.7\Ms$. The resulting 2g-IMBH receives a GW kick with amplitude $v_{\rm kick} > 99\kms$ at 95\% confidence level. The probability to obtain a kick of $\simeq 50 \kms$ is of the order of $\sim 0.1\%$, regardless of the spin distribution choice. 
Therefore, we re-run the simulation shortly before the merger event and assign to the merger remnant either a small ($v_\gw = 20\kms$) or large ($v_\gw = 100\kms$) recoil kick. In the case of $v_{\rm rec}=100\kms$ the merger remnant promptly leaves the cluster, as expected. 

In the case of $v_\gw=20\kms$, instead, the 2g-IMBH remains in the cluster core and undergoes a series of resonant interactions with two BHs, which drives the IMBH to merge after just $25.5$ Myr with an upper-mass gap BH with ($m_{\rm BH,2} = 63\Ms$). The 3g-IMBH, with a mass $m_{\rm 3g}\simeq 300\Ms$, receives a kick $v_{\rm GW} > 90\kms$ regardless of the amplitude and direction of progenitors' spins, hence it leaves the cluster promptly after the merging event. 

The impact of relativistic effects on the chaotic nature of $N$-body dynamics is apparent in this case: The displacement caused by the GW recoil favor the onset of the three-body interactions that led to the merger. For comparison, in the zero-kick model the two BHs never find each other.  

\section{Conclusion}
\label{sec:end}

In this work we have analysed the properties of IMBHs formed in the \dragonii cluster models, a suite of 19 direct $N$-body simulations representing star clusters initially made up of $\leq 10^6$ stars, up to $33\%$ of which initially paired in a binary. Our main results can be summarised as follows:
\begin{itemize}    
    \item Out of 19 models, 8 IMBHs form in \dragonii clusters, following three main formation channels: a) collapse of a VMS formed via repeated stellar mergers (2 IMBHs), b) accretion of stellar material onto stellar BHs (1), c) BH-BH mergers (5). The IMBHs have typical masses in the range $m_{\rm IMBH} = (100-370)\Ms$. Aside IMBH seeding, the aforementioned formation channels significantly contribute to the population of BHs with masses in the upper mass-gap, for which we derive a formation efficiency of $\eta_{\rm gap} = 3.44\times 10^{-5}\Ms^{-1}$ [Table \ref{tab:t1} and Figures \ref{fig:fig3a}-\ref{fig:gap}].
    \item Despite the small sample, we find a striking relation between the IMBH formation channel and the host cluster properties. Stellar mergers dominate IMBH formation in the densest clusters, operating on short timescale ($10$ Myr) and producing the most massive IMBHs ($>200\Ms$). Star-BH interactions and BBH mergers, instead, dominate IMBH formation in less dense clusters, showing that the looser the cluster the longer the IMBH formation time ($10-300$ Myr), and the larger the IMBH seed mass [Figure \ref{fig:imbF}].
    \item When relativistic recoil is neglected, Newtonian dynamics represents a serious threat to IMBH retention and growth. In fact, all IMBHs are ejected from \dragonii cluster through strong dynamical interactions. Nonetheless, in the Newtonian scenario some IMBHs undergo multiple IMBH-BH mergers reaching up to the fourth generation. The inclusion of GW recoil severely impacts the IMBH growth process, limiting the IMBH merger history to two generations. We implement a simple model for BH natal spins, based on stellar evolution models, to infer the IMBH mass and spins. In our fiducial model IMBHs are characterised by masses up to $376\Ms$ and relatively large spins, i.e. $\chi_{\rm IMBH} > 0.6$. The inclusion of relativistic kicks in the simulations enables a fully self-consistent description of the IMBH merging process and reveal how hard is for IMBHs to be retained in their parent clusters. Nonetheless, even in the unlikely case the IMBH receives small GW kicks and avoid ejection, our simulations confirm how chaotic and unpredictable the evolution of the post-merger IMBH can be. For example, in one simulation the inclusion of the kick can favour the merger of the IMBH with a BH more massive than in the zero GW kick case [Table \ref{tab:tabimbh} and Figure \ref{fig:IBHs}]. 
\end{itemize}

The \dragonii simulations represent one of the few numerical models \citep[see also e.g.][]{2022MNRAS.511.5797M,2022MNRAS.514.5879M} in which all the three main channels proposed for the formation of IMBHs have been confirmed. Our analysis of the \dragonii database suggests that: i) IMBHs form preferentially via collapse of stellar merger products (BBH mergers) in clusters more (less) dense than $3\times10^5\Ms$ pc$^{-3}$, ii) have large spins at formation $\chi_{\rm BH} > 0.6$, iii) live most of their life with a BH companion, iv) are unlikely to grow beyond a few hundred $\Ms$ because of the efficiency of dynamical scatterings and the impact of relativistic recoil.

\appendix
\section{The evolution and growth of IMBHs in \textsc{Dragon-II} clusters}
\label{sec:IMevo}

In this section, we discuss in detail the evolutionary history of the 8 IMBHs in \dragonii clusters, their main properties, and retention probability. In the following we indicate with BH$1,~2$ and with letters $a,~b$ the IMBH progenitors, and with $p1,~p2$ the progenitors of the IMBH progenitors, in such a way that $p1a, ~p2a$ indicates the two progenitors of the primary BH that eventually led to the IMBH. 
All the main properties of the IMBHs are summarised in Table \ref{tab:tabimbh}.

\paragraph*{IMBH No. 1: IBH\_Rh1.75f5N1000k.}
In one cluster model with $R_\ham=1.75$ pc, $f_b=0.05$, $N=10^6$, the IMBH forms via the merger of two BHs with masses $m_{\rm BH,1} = 86.3\Ms$ and $m_{\rm BH,2} = 58.9\Ms$. The primary BH is the byproduct of a merger between a PPISN BH and a massive star in the HG phase $m_{p1a}+m_{p2a} = (40.5 + 91.7)\Ms$ in a primordial binary, and we assume that it spins-up during its growth, assigning it a spin $\chi_{\rm BH,1} > 0.8$. The secondary BH, instead, forms from the merger of two stars in a primordial mass, with masses $m_{p1b}+m_{p2b} = (37+82)\Ms$, with the lighter component being a naked He MS star and the heavier a star in the HG phase. We assign the companion BH a spin $\chi_{BH,2} = 0.01$. 
The resulting IMBH (2g) has a mass $m_{\rm 2g} = 138.4^{+1.8}_{-3.0}\Ms$ and spin $\chi_{\rm 2g} = 0.76^{+0.11}_{-0.27}$, with the spin increasing at decreasing the mass. In the simulation with GW recoil disabled, the IMBH forms a binary with a BH with mass $m_{\rm BH} = 40.5\Ms$ --- formed from a single star --- and ultimately merge after being ejected outside the cluster, leading to a final IMBH (3g) with a mass $m_{\rm 3g} = 174.0^{+2.6}_{-4.6}\Ms$ and $\chi_{\rm 3g}=0.68^{+0.20}_{-0.40}$. However, the GW recoil associated with the formation of the 2g-IMBH is sufficiently large ($v_{\rm GW} = 150-2200\kms$) to make the retention of the IMBH and its further growth impossible. 

\paragraph*{IMBH No. 2: IBH\_Rh1.75f20N120k.}
The second IMBH in the sample (simulation with $R_\ham=1.75$ pc, $f_b=0.2$, $N=120$) forms through a BH-BH merger with component masses $m_{\rm BH,1} + m_{\rm BH,2} = (95.5+95.8)\Ms$. The previous evolution of these massive BHs is rather complex. The primary forms from the accretion of a MS star with mass $m_{\rm p2a}= 110\Ms$ and a BH ($m_{\rm p1a}=40.5\Ms$) previously formed from the merger of two MS stars in a primordial binary. We thus assign the primary BH a spin $\chi_{\rm BH,1}=0.8-1$. The secondary, instead, forms from the merging of two stars in a primordial binary during the HG phase of the heavier component. We assign the secondary BH a small spin $\chi_{\rm BH,2} = 0.01$. The resulting IMBH (2g) has a mass $m_{\rm 2g}=181.8^{+1.8}_{-2.7}\Ms$ and spin $\chi_{\rm 2g} = 0.72^{+0.10}_{-0.15}$. When GW recoil is disabled, the IMBH undergoes a second merger with a BH with mass $m_{\rm BH,2} = 40.5\Ms$ that did not experience significant mass-transfer, thus likely characterised by a low spin. After the merger, the IMBH (3g) has a mass $m_{\rm 3g} = 217.8^{+2.5}_{-4.3}\Ms$ and spin $\chi_{\rm 3g} = 0.65^{+0.20}_{-0.45}$. It forms a binary that is ejected and merges outside the cluster, leaving a 4g-IMBH with final mass $m_{\rm 4g} = 253.9^{+2.9}_{-5.9}\Ms$ and spin $\chi_{\rm 4g} = 0.56^{+0.28}_{-0.34}$.
There is a probability of $\sim 0.2\%$ for the GW recoil imparted on the 2g-IMBH to remain below $v_{\gw} < 20\kms$, i.e. sufficiently smaller to be retained in the cluster. However, when the 3g-IMBH forms, the post-merger kick is in the range $v_{\rm GW} = 35-2000\kms$, definitely larger than the cluster escape velocity. We discuss the results from a self-consistent simulation of the evolution of the 2g-IMBH in Section \ref{sec:self}. 

\paragraph*{IMBH No. 3: IBH\_Rh1.75f20N600k.}
The third IMBH forms in model with $R_\ham = 1.75$ pc, $f_b=0.2$, and $N=600,000$ through the merger of two BHs with mass $m_{\rm BH,1}=74.7\Ms$ and $m_{\rm BH,2} = 68.8\Ms$, both being byproduct of a stellar merger event in two primordial binaries. We assume that both BHs have negligible spins, which leads to an IMBH (2g) with a mass $m_{\rm 2g} = 136.6^{+1.2}_{-1.9}$ and spin $\chi_{\rm 2g} = 0.72^{+0.08}_{-0.15}$. The post-merger recoil is sufficiently small ($v_{\rm GW} = 20-45\kms$) to retain the IMBH. The IMBH eventually merges with a BH with mass $m_{\rm BH,2} = 18\Ms$ (for which $\chi_{\rm BH,2} = 0.01$) after being ejected from the cluster. The final IMBH (3g) has a mass $m_{\rm 3g} = 152.7^{+1.5}_{-2.4}\Ms$ and spin $\chi_{\rm 3g}=0.61^{+0.22}_{-0.36}$.
 
\paragraph*{IMBH No. 4: IBH\_Rh0.8f20N120k.}
The fourth IMBH forms in model $R_\ham = 0.8$ pc, $f_b = 0.2$, $N=120,000$ from two BHs with masses $m_{\rm BH,1} = 79.8\Ms$ and $m_{\rm BH,2}=40.5\Ms$. The primary formed from a star-BH merger in a primordial binary involving a BH $m_{\rm p1a} = 40.5\Ms$ and a star in the HG phase with mass $m_{\rm p2a} = 78.5$. We assign a spin $\chi_{\rm BH,1} > 0.8$ to the primary and a small spin to the secondary, which did not undergo any significant matter accretion phase. The IMBH (2g) formed this way has a mass $m_{\rm 2g} = 115.6^{+1.3}_{-3.0}\Ms$ and spin $\chi_{\rm 2g} = 0.74^{+0.15}_{-0.36}$. In absence of GW recoil, the IMBH captures a BH with mass $m_{\rm BH,2 }=39\Ms$, which experienced mass transfer in a primordial binary, and finally merge outside the cluster. In this case, we assign to the stellar BH a spin in the $0-1$ range, which leads to an IMBH (3g) with final mass $m_{\rm 3g}=149.8^{+2.0}_{-4.6}\Ms$ and $\chi_{\rm 3g}=0.67^{+0.22}_{-0.35}$. The kick received by the 2g-IMBH, however, is large enough ($v_{\rm GW} > 100\kms$) to kick the IMBH out before the binary can form.

\paragraph*{IMBH No. 5: IBH\_Rh0.8f20N120k.}
Even the fifth IMBH, which forms in model $R_\ham=0.8$ pc, $f_b=0.2$, and $N=120,000$, is the byproduct of a BBH merger. The primary, with a mass $m_{\rm BH,1}=80.7\Ms$, forms from the merger of two MS stars, and we assume negligible spin. The companion, with a mass $m_{\rm BH,2}=51.5\Ms$, forms from mass transfer in a primordial binary, thus we assume that its spin is distributed in the $\chi_{\rm BH,2} = 0.8-1$ range. The resulting IMBH has a mass $m_{\rm 2g} = 126.4^{+0.7}_{-1.0}\Ms$ and spin $\chi_{\rm 2g} = 0.67^{+0.06}_{-0.08}$.  In the case of no GW recoil, the IMBH captures a BH with mass $m_{\rm BH} = 30\Ms$ formed from a single star (thus $\chi_{\rm BH} = 0.01$), and the resulting binary is eventually ejected from the cluster, ultimately merging outside the cluster and leaving behind an IMBH with mass $m_{\rm 3g} = 153.0^{+1.4}_{-2.1}\Ms$ and spin $\chi_{\rm 3g} = 0.62^{+0.19}_{-0.42}$. Even in this case, though, the GW kick imparted onto the 2g-IMBH ($v_{\rm GW} > 60\kms$) is larger than the cluster escape velocity. 

\paragraph*{IMBH No. 6: IBH\_Rh0.8f20N300k.}
The sixth IMBH forms in a cluster with $R_\ham=0.8$ pc, $f_b=0.2$, and $N=300,000$, from the coalescence of a PPISN BH ($m_{\rm BH} = 40.5\Ms$, negligible spin) and a massive star in the HG phase ($m_{\rm HG}=133\Ms$). The IMBH, with mass $m_{\rm 1g} = 107\Ms$, likely spins-up during the interaction with its stellar companion. The IMBH is eventually ejected as a single object in consequence of a resonant strong scattering involving two BHs with masses $m_{\rm BH,1} = 35.2\Ms$ and $m_{\rm BH,2} = 67.7\Ms$. 

\paragraph*{IMBH No. 7: IBH\_Rh0.47f20N120k.}
The seventh, and most massive, IMBH, forms in one of the most compact \dragonii clusters ($R_\ham=0.47$ pc, $f_b=0.2$, and $N=120,000$). A complex series of stellar mergers triggers the IMBH seeding, leading to an IMBH with mass $m_{\rm 1g} = 288\Ms$ that eventually collides with a massive MS star with mass $m_{\rm MS} = 122\Ms$. The resulting IMBH, which can be considered half-way between first and second generation, has a mass $m_{\rm 1g*} = 350\Ms$ and likely a large spin, $\chi_{\rm 1g*} \sim 0.8-1$, owing to the mass accretion process. The IMBH captures a stellar BH with mass $m_{\rm BH,2} = 29\Ms$ formed from a single star, for which we assume negligible spin. The IMBH-BH binary is eventually ejected in a strong binary-single interaction and merges outside the cluster, leading to a 2g-IMBH with mass $m_{\rm 2g} = 376.5^{+0.8}_{-3.7}\Ms$ and spin $\chi_{\rm 2g} = 0.79^{+0.17}_{-0.27}$.

\paragraph*{IMBH No. 8: IBH\_Rh0.47f20N300k.}
The last IMBH forms in the densest \dragonii cluster ($R_\ham=0.47$ pc, $f_b=0.2$, and $N=300,000$). Initially, an IMBH seed with mass $m_{\rm 1g} = 189 \Ms$ forms via subsequent mergers of massive stars. It later collides with a MS star with mass $m_{\rm MS} = 51.7$ and shortly after with two low mass stars, leaving behind an IMBH (1g*) with mass $m_{\rm 1g*} = 217\Ms$ and high-spin triggered by mass accretion. The IMBH undergoes merger with a low-spin BH with mass $m_{\rm BH} = 27\Ms$, forming a 2g-IMBH with a mass $m_{\rm 2g} = 241.4^{+0.8}_{-3.3}\Ms$ and spin $\chi_{\rm 2g} = 0.77^{+0.18}_{-0.37}$. 
In absence of GW recoil, the 2g-IMBH further merge with a low-spin BH (mass $m_{\rm BH} = 38\Ms$) after being ejected in the cluster, leading to a 3g-IMBH characterised by $m_{\rm 3g} = 275.3^{+1.8}_{-5.4}\Ms$ and spin $\chi_{\rm 3g} = 0.63^{+0.28}_{-0.39}$. When GW recoil are taken into account, the 2g-IMBH receives a kick $v_{\rm GW} > 40\kms$, thus larger than the cluster escape velocity. We explore more in detail the retention of this IMBH in Section \ref{sec:reten}.

\section*{Acknowledgements}
The authors thank the referee for their constructive report and feedback. The authors warmly thank Agostino Leveque for their help and assistance in using their implementation of the \mcl code, and Giuliano Iorio, Sara Rastello, and Michela Mapelli for useful comments and discussion. 

This work benefited of the support from the Volkswagen Foundation Trilateral Partnership through project No.~97778 ``Dynamical Mechanisms of Accretion in Galactic Nuclei'' and the Deutsche Forschungsgemeinschaft (DFG, German Research Foundation) -- Project-ID 138713538 -- SFB 881 ``The Milky Way System''), and by the COST Action CA16104 ``GWverse''. The authors gratefully acknowledge the Gauss Centre for Supercomputing e.V. for funding this project by providing computing time through the John von Neumann Institute for Computing (NIC) on the GCS Supercomputer JUWELS Booster at Jülich Supercomputing Centre (JSC).

MAS acknowledges funding from the European Union’s Horizon 2020 research and innovation programme under the Marie Skłodowska-Curie grant agreement No.~101025436 (project GRACE-BH, PI: Manuel Arca Sedda). 

AWHK is a fellow of the International Max Planck Research School for Astronomy and Cosmic Physics at the University of Heidelberg (IMPRS-HD).
The work of PB was supported by the Volkswagen Foundation under the special stipend No.~9B870.

PB acknowledge the support within the grant No.~AP14869395 of the Science Committee of the Ministry of Science and Higher Education of Kazakhstan ("Triune model of Galactic center dynamical evolution on cosmological time scale"). 
The work of PB was supported under the special program of the NRF of Ukraine Leading and Young Scientists Research Support - "Astrophysical Relativistic Galactic Objects (ARGO): life cycle of active nucleus", No.~2020.02/0346.

RS acknowledges support by Yunnan Academician Workstation of Wang Jingxiu (No. 202005AF150025) and thanks Max Planck Institute for Astrophysics (Thorsten Naab) for hospitality during many visits.

MG was partially supported by the Polish National Science Center (NCN) through the grant No. 2021/41/B/ST9/01191.

FPR acknowledge the support by the European Research Council via ERC Consolidator Grant KETJU (no. 818930).

\section*{Data Availability}
The data from the runs of these simulations and their initial models
will be made available upon reasonable request by the corresponding author. 
The \textsc{Nbody6++GPU} code is publicly available\footnote{\url{https://github.com/nbody6ppgpu/Nbody6PPGPU-beijing}}. The \textsc{McLuster} version used in this work will soon be available. A similar version is described in \cite{2022MNRAS.514.5739L}.


\bibliographystyle{mnras}
\bibliography{example} 

\bsp	
\label{lastpage}
\end{document}